\documentclass[twocolumn,floatfix,amsmath,amssymb,superscriptaddress,nofootinbib,longbibliography]{revtex4-2}

\usepackage{lipsum}
\usepackage{verbatim}
\usepackage[colorlinks=true,linkcolor=blue,citecolor=blue]{hyperref}
\usepackage[utf8]{inputenc}
\usepackage[T1]{fontenc}
\usepackage{xspace}
\usepackage{enumitem}

\usepackage{array}
\usepackage{mathtools}
\usepackage[mathscr]{euscript}

\usepackage{physics}
\usepackage{algorithmic}
\usepackage{algorithm}
\usepackage{siunitx}

\usepackage{graphicx}
\usepackage{tikz}
\usepackage{pgfplots}

\usepackage{tabularx}
\usepackage{booktabs}
\usepackage{multirow}

\newcolumntype{x}[1]{>{\centering\arraybackslash\hspace{0pt}}p{#1}}

\DeclarePairedDelimiter\ceil{\lceil}{\rceil}

\DeclareMathOperator*{\argmax}{argmax}
\DeclareMathOperator*{\argmin}{argmin}
\DeclareMathOperator*{\poly}{poly}

\usepackage[caption=false]{subfig}
\newcommand{\phantomsubfloat}[1]{
    {%
        \captionsetup[subfigure]{labelformat=empty}
        \subfloat[][]{#1}
    }%
}

\usepackage{cleveref}
\crefname{equation}{Eq.}{Eqs.}
\crefname{section}{Sec.}{Secs.}
\crefname{subsection}{Sec.}{Secs.}
\crefname{appendix}{Appendix}{Appendices}
\crefname{figure}{Figure}{Figures}
\crefname{table}{Table}{Tables}
\newcommand{\methods}{\hyperref[sec:methods]{Methods}\xspace}
\newcommand{\methodsabove}[1]{\hyperref[#1]{above}\xspace}
\newcommand{\methodsbelow}[1]{\hyperref[#1]{below}\xspace}
\newcommand{\inertlink}{\textcolor{blue}}

\newcommand{\ie}{\textit{i}.\textit{e}.}
\newcommand{\eg}{\textit{e}.\textit{g}.}

\usepackage{xcolor}
\definecolor{daxcolor}{rgb}{0.8, 0.3, 0.4}

\usepackage{orcidlink}

\usepackage[final]{pdfpages}
\makeatletter
\AtBeginDocument{\let\LS@rot\@undefined}
\makeatother

\begin{document}

\preprint{}
\title{Interacting Non-Hermitian Edge and Cluster Bursts on a Digital Quantum Processor}

\author{Jin Ming Koh\,\orcidlink{0000-0002-6130-5591}}
\affiliation{Department of Physics, Harvard University, Cambridge, Massachusetts 02138, USA}
\affiliation{Quantum Innovation Centre (Q.InC), Agency for Science, Technology and Research (A*STAR), 2 Fusionopolis Way, Innovis \#08-03, Singapore 138634, Republic of Singapore\looseness=-1}

\author{Wen-Tan Xue\,\orcidlink{0000-0001-7823-9888}}
\affiliation{Department of Physics, National University of Singapore, Singapore 117542, Republic of Singapore}

\author{Tommy Tai\,\orcidlink{0000-0002-9478-5499}}
\affiliation{Department of Physics, Massachusetts Institute of Technology, Cambridge, Massachusetts 02142, USA}

\author{Dax Enshan Koh\,\orcidlink{0000-0002-8968-591X}}
\affiliation{Quantum Innovation Centre (Q.InC), Agency for Science, Technology and Research (A*STAR), 2 Fusionopolis Way, Innovis \#08-03, Singapore 138634, Republic of Singapore\looseness=-1}
\affiliation{Institute of High Performance Computing (IHPC), Agency for Science, Technology and Research (A*STAR), 1 Fusionopolis Way, \#16-16 Connexis, Singapore 138632, Republic of Singapore\looseness=-1}
\affiliation{Science, Mathematics and Technology Cluster, Singapore University of Technology and Design, 8 Somapah Road, Singapore 487372, Republic of Singapore\looseness=-1}

\author{Ching Hua Lee\,\orcidlink{0000-0003-0690-3238}}
\email{phylch@nus.edu.sg}
\affiliation{Department of Physics, National University of Singapore, Singapore 117542, Republic of Singapore}

\begin{abstract}
A lossy quantum system harboring the non-Hermitian skin effect can in certain conditions exhibit anomalously high loss at the boundaries of the system compared to the bulk, a phenomenon termed the non-Hermitian edge burst. We uncover interacting many-body extensions of the edge burst that are spatially extended and patterned, as well as cluster bursts that occur away from boundaries. Owing to the methodological difficulty and overhead of accurately realizing non-Hermitian dynamical evolution, much less tunable interactions, few experimental avenues in studying the single-particle edge burst have been reported to date and none for many-body variants. We overcome these roadblocks in this study, and present a realization of edge and cluster bursts in an interacting quantum ladder model on a superconducting quantum processor. We utilize a time-stepping algorithm, which implements time-evolution by non-Hermitian Hamiltonians by composing a linear combination of unitaries scheme and product formulae, to assess long-time behavior of the system. We observe signatures of the non-Hermitian edge burst on up to $64$ unit cells, and detect the closing of the dissipative gap, a necessary condition for the edge burst, by probing the imaginary spectrum of the system. In suitable interacting regimes, we identify the emergence of spatial patterning and cluster bursts. Beyond establishing these generalized forms of edge burst phenomena, our study paves the way for digital quantum processors to be harnessed as a versatile platform for non-Hermitian condensed-matter physics.
\end{abstract}

\maketitle
\date{\today}

\section{Introduction}
\label{sec:intro}

\begin{figure*}[!t]
    \centering
    \includegraphics[width = \linewidth]{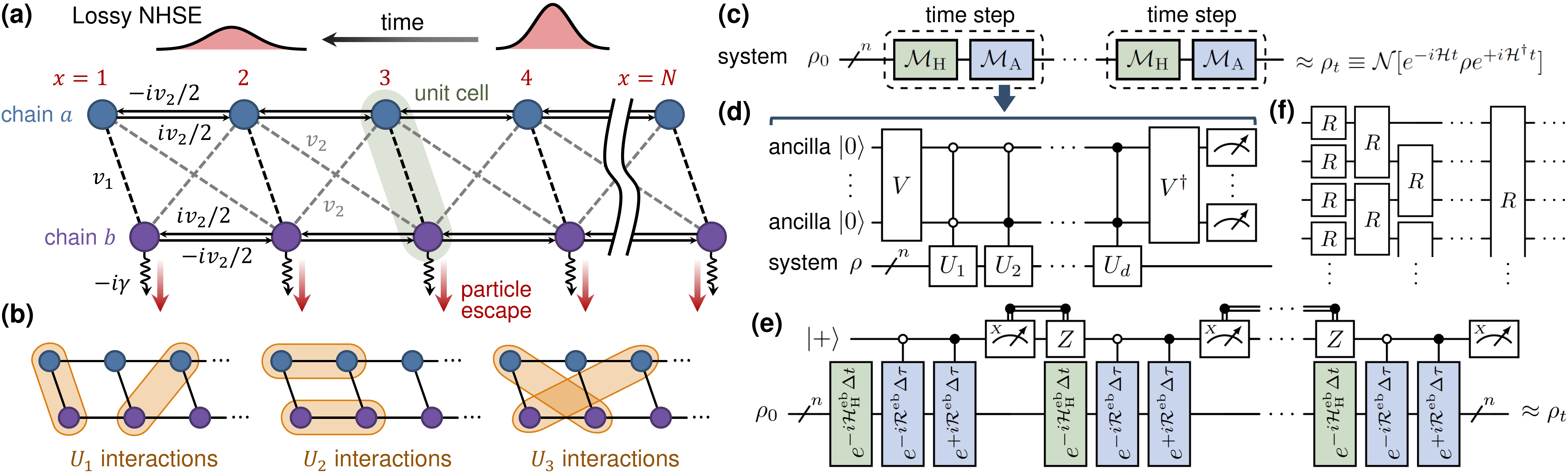}
    \phantomsubfloat{\label{fig:schematic/schematic-model-non-interacting}}
    \phantomsubfloat{\label{fig:schematic/schematic-model-interactions}}
    \phantomsubfloat{\label{fig:schematic/schematic-circuit-time-stepping}}
    \phantomsubfloat{\label{fig:schematic/schematic-circuit-lcu}}
    \phantomsubfloat{\label{fig:schematic/schematic-circuit-experiment}}
    \phantomsubfloat{\label{fig:schematic/schematic-circuit-trotterization}}
    \vspace{-2\baselineskip}
    \caption{\textbf{Non-Hermitian interacting quantum ladder model and quantum simulation methodology.} \textbf{(a)} Schematic of quantum ladder built from sublattice chains $a$ and $b$, with reciprocal hopping $v_1, v_2$ between sites and on-site loss $\gamma$ on the $b$ chain. Leftward drift of a wavepacket is induced by the non-Hermitian skin effect (NHSE), accompanied by dampening of wavepacket amplitude by the non-Hermitian loss. \textbf{(b)} A sequence of range-$r$ density-density interactions $\smash{\{U_r\}_r}$ between sites on the ladder, giving rise to many-body effects enriching the non-Hermitian edge burst. \textbf{(c)} Structure of time-stepping quantum algorithm simulating time-evolution by a general non-Hermitian Hamiltonian on a quantum processor. The quantum maps $\mathcal{M}_{\mathrm{H}}$ and $\mathcal{M}_{\mathrm{A}}$ perform time-evolution by the Hermitian and anti-Hermitian parts of the Hamiltonian respectively. \textbf{(d)} The non-unitary $\mathcal{M}_{\mathrm{A}}$ is implemented through a linear combination of unitaries (LCU) circuit primitive, with the aid of a small ancillary register. \textbf{(e)} Structure of time-evolution circuits executed on the quantum processor. The LCU implementing $\mathcal{M}_{\mathrm{A}}$ superposes forward and backward time-evolution, coherently controlled by a single ancillary qubit that is re-used through mid-circuit qubit reset. Observables such as site-resolved occupancies are measured on the time-evolved state at the end of the circuit. \textbf{(f)} Time-evolution by Hermitian Hamiltonian components by trotterization. The $R$ gates denote varying multi-qubit Pauli rotations.}
    \label{fig:schematic}
\end{figure*}

While closed quantum-mechanical systems conserve energy and are described by Hermitian Hamiltonians, open quantum systems lead to the possibility of non-Hermitian physical descriptions~\cite{ashida2020non, okuma2023non, lin2023topological, zhang2022review}. Such non-Hermitian systems typically feature effective gain or loss coupled to an external environment. Canonical examples of this broader category of quantum systems span lossy optics~\cite{el2019dawn}, quantum electrodynamics circuits~\cite{blais2021circuit}, and electronic response in certain materials~\cite{ding2022non}. Amongst other applications, the properties endowed by non-Hermiticity have been harnessed to great effect in the engineering of ultra-sensitive sensors and detectors~\cite{budich2020non, wiersig2020prospects, edvardsson2022sensitivity}.

Indeed, non-Hermiticity enables unique physics with no Hermitian analog. A paradigmatic example is the emergence of the non-Hermitian skin effect (NHSE) \cite{yao2018edge,Lee2019,Youkomizo2019,Longhi2019,Okuma2020}, wherein a quantum system exhibits an asymmetric probability current flow in its bulk, an extensive number of eigenstates localized at a boundary, and profound differences in spectrum and dynamics dependent on boundary conditions---properties that are in contrast to conventional Hermitian systems. While the NHSE and generalizations~\cite{Zhang2022, shen2022, Li2020critical, Mu2020, yang2022designing} are well-studied, recent discoveries uncovered a new phenomenon labeled the non-Hermitian edge burst~\cite{xue2022non,hu2023steady,Wang_2021,yuce2023non,wen2024investigation} occurring in settings with dissipative (\ie~imaginary energy) gap closure. In non-Hermitian lossy quantum systems, the edge burst is characterized by anomalously high particle leakage near a boundary, supported by novel algebraic long-ranged decay of wavefunction amplitudes in the bulk. Beyond the canonical single-particle setting~\cite{xiao2024observation,zhu2024observation}, the interplay of the edge burst with interacting many-body physics is uncharted terrain in both theory and experiments. 

Here, we report physical realizations of both single-body and novel many-body interacting generalizations of the edge burst on superconducting transmon-based quantum devices. We develop an efficient digital quantum simulation methodology for general non-Hermitian Hamiltonians, which leverages a linear combination of unitaries circuit construction technique in a time-stepping algorithm for dynamical evolution, and apply our method to probe a minimal quantum ladder system supporting the edge burst. Beyond observing signatures of the conventional single-body non-interacting edge burst on up to $64$ unit cells, we show on quantum hardware that introducing sequences of density-density interactions lead to the formation of spatially extended and ordered (\ie~patterned) versions of the edge burst, and in certain regimes can also lead to bursts occurring in the bulk of the system far from boundaries, which we dub cluster bursts. 
 
Our use of a quantum simulator enabled versatile accommodation of interactions of tunable strength and range in a model hosting multiple particles, an important advantage over alternative effectively single-body platforms such as waveguide photonics~\cite{Li2023_ep} and analog circuits~\cite{Helbig2020}. While quantum processors are increasingly utilized for quantum dynamics~\cite{rubin2024quantum, google2023measurement, koh2023measurement, jafferis2022traversable, daley2022practical, tan2021domain, sun2024quantum} and various areas of (Hermitian) condensed-matter applications~\cite{mi2022time, clinton2024towards, ebadi2021quantum, kim2023evidence, satzinger2021realizing, google2023non, koh2024realization, koh2022simulation, koh2022stabilizing}, their use in studying non-Hermitian physics has remained nascent owing to difficulties in scalably realizing non-Hermitian time-evolution on near-term noisy intermediate-scale quantum (NISQ) devices of the present. Demonstrations to-date are restricted to small system sizes or rely on resource intensive protocols~\cite{jebraeilli2025quantum, bian2023quantum, wen2019experimental, shen2025observation, shen2024enhanced, liu2023practical}. In addition to unveiling novel forms of the edge burst induced by interactions, the methods we develop in this work, which do not require expensive classical pre-processing of Hamiltonian matrices nor large numbers of intermediary measurements or variational iterations, bring us significantly closer to meaningfully utilizing quantum processors as a platform for studying generic non-Hermitian systems.

\section{Results}
\label{sec:results}

\subsection{Minimal interacting non-Hermitian quantum ladder}
\label{sec:results/model}

While the quantum simulation methods we utilize are general, we focus on investigating the phenomenology of the non-Hermitian edge burst in the present work. To set a clear physical picture, we first establish the theoretical setting underlying our study. The model we investigate is built atop a minimal bosonic one-dimensional quantum ladder exhibiting the edge burst~\cite{xue2022non}, characterized by a two-band Bloch Hamiltonian
\begin{equation}\begin{split}
    \mathcal{H}^{\mathrm{eb}}_0(k) &= \left( v_1 + v_2 \cos{k} \right) \sigma^x \\
        &\qquad
        + \left( v_2 \sin{k} + \frac{i \gamma}{2} \right) \sigma^z
        - \frac{i \gamma}{2} \mathbb{I},
\end{split}\end{equation}
where $v_1, v_2 \geq 0$ are tight-binding hopping coefficients, $\gamma > 0$ is an on-site loss rate and is responsible for the non-Hermiticity of the system, and $\sigma^x, \sigma^z$ are Pauli operators acting on two pseudospin degrees of freedom which we associate with $a$ and $b$ sublattices within a unit cell.

We illustrate $\mathcal{H}^{\mathrm{eb}}_0$, which is interpretable as a lossy quantum walk Hamiltonian, in real space in \cref{fig:schematic/schematic-model-non-interacting}. The $\gamma$ loss corresponds physically to leakage or escape~\cite{xue2022non} of the quantum walker into the environment. As long as $v_1 \neq 0$, the $\pi / 2$ fluxes in the triangles between the $a$- and $b$-sublattice chains generate rotational motion such that they favor opposing directions of travel along the ladder; but the $\gamma$ loss on $b$ sublattices dampens dynamics on that chain, thereby producing a preferential leftward chiral motion (in the $-\vu{x}$ direction). This mechanism gives rise to the NHSE along the ladder, which localizes an extensive number of bulk-band eigenstates on the left boundary ($x = 1$) of the system. Alternatively, the NHSE can be understood as a consequence of an equivalence~\cite{xue2022non, wen2024investigation} of $\mathcal{H}^{\mathrm{eb}}_0$ to the non-Hermitian Su-Schrieffer-Heeger (SSH) model with asymmetric left-right particle hoppings~\cite{yao2018edge}.

Here, we extend the edge burst phenomenon from the canonical non-interacting single-particle context to a many-body interacting setting. In addition to $\mathcal{H}^{\mathrm{eb}}_0$, we consider also a natural sequence of density-density interactions (see \cref{fig:schematic/schematic-model-interactions}). In real space over $N$ unit cells, the model we examine is $\mathcal{H}^{\mathrm{eb}} = \mathcal{H}^{\mathrm{eb}}_0 + \mathcal{H}^{\mathrm{eb}}_{\mathrm{int}}$, where
\begin{equation}\begin{split}
    \mathcal{H}^{\mathrm{eb}}_{\mathrm{int}}
    = \sum_{r \geq 1} U_r \sum_{z = 1}^{2 N} 
            n_z n_{z + r},
\end{split}\end{equation}
where $U_r \in \mathbb{R}$ are range-$r$ interaction strengths and $n_z$ is the number operator on site $z$ on the flattened ladder, that is, $n_z = n_{x \ell} = \smash{c^\dag_{x \ell} c_{x \ell}}$ for flattened index $z = 2 x - \delta_{\ell a}$ and particle operator $c_{x \ell}$ acting on unit cell $x \in [N]$ and sublattice $\ell \in \{a, b\}$. Physically, $U_r$ introduces an interaction potential dependent on the occupation of sites on the ladder separated by distance $r$. We consider hardcore bosons on the ladder, which emerge in a regime of strong on-site repulsive interactions on top of $\mathcal{H}^{\mathrm{eb}}_{\mathrm{int}}$ (see \methods).

The Schrödinger equation $\smash{i (\dd / \dd t) \omega_t = \comm*{\mathcal{H}^{\mathrm{eb}}}{\omega_t}}$ prescribes that from an initial normalized quantum state (density matrix) $\omega_0$ of the ladder, the state at time $t$ is given by $\smash{\omega_t = V^{\mathrm{eb}}_t \omega_0 (V^{\mathrm{eb}}_t)^\dag}$ where $\smash{V^{\mathrm{eb}}_t = \exp(-i \mathcal{H}^{\mathrm{eb}} t)}$ is the time-evolution propagator. As $\mathcal{H}^{\mathrm{eb}}$ is non-Hermitian, the propagator $\smash{V^{\mathrm{eb}}_t}$ is non-unitary and state normalization is not generically preserved. In particular, the quantum state norm decays as 
\begin{equation}\begin{split}
    \dv{t} \tr(\omega_t)
    = i \expval{{\mathcal{H}^{\mathrm{eb}}}^\dag - \mathcal{H}^{\mathrm{eb}}}_{\omega_t}
    \hspace{-3mm} 
    = -2 \gamma \sum_{x = 1}^N \expval{n_{x b}}_{\omega_t}
    \leq 0.
    \label{eq:results/model/wavefunction-norm}
\end{split}\end{equation}

The norm of the quantum state describes the probability of the walker (\ie~boson) remaining on the ladder. Accordingly, the escape probability of the walker at unit cell $x$ by time $t$ is given by the time integral
\begin{equation}\begin{split}
    P_x(t) 
    = 2 \gamma \int_0^t \expval{n_{x b}}_{\omega_\tau} \, \dd{\tau},
    \label{eq:results/model/escape-probs}
\end{split}\end{equation}
and the final cell-resolved escape probabilities are given by the long-time limit $\mathcal{P}_x = \lim_{t \to \infty} P_x(t)$. Restricted to pure single-particle quantum states, \cref{eq:results/model/wavefunction-norm,eq:results/model/escape-probs} reduce to be consistent with expressions in Ref.~\cite{xue2022non}.

Our central quantities of interest are the final escape probabilities $\mathcal{P}_x$ which enable direct probing of the presence and properties of the non-Hermitian edge burst. Accordingly, we seek to measure the site-resolved occupancy densities $\smash{\expval{n_{x \ell}}_{\omega_t}}$ in our realizations, from which the escape probabilities $P_x(t)$ and $\mathcal{P}_x$ can be recovered.

\subsection{Realizing non-Hermitian quantum dynamics on quantum hardware}
\label{sec:results/simulation}

\begin{figure*}
    \centering
    \includegraphics[width = \linewidth]{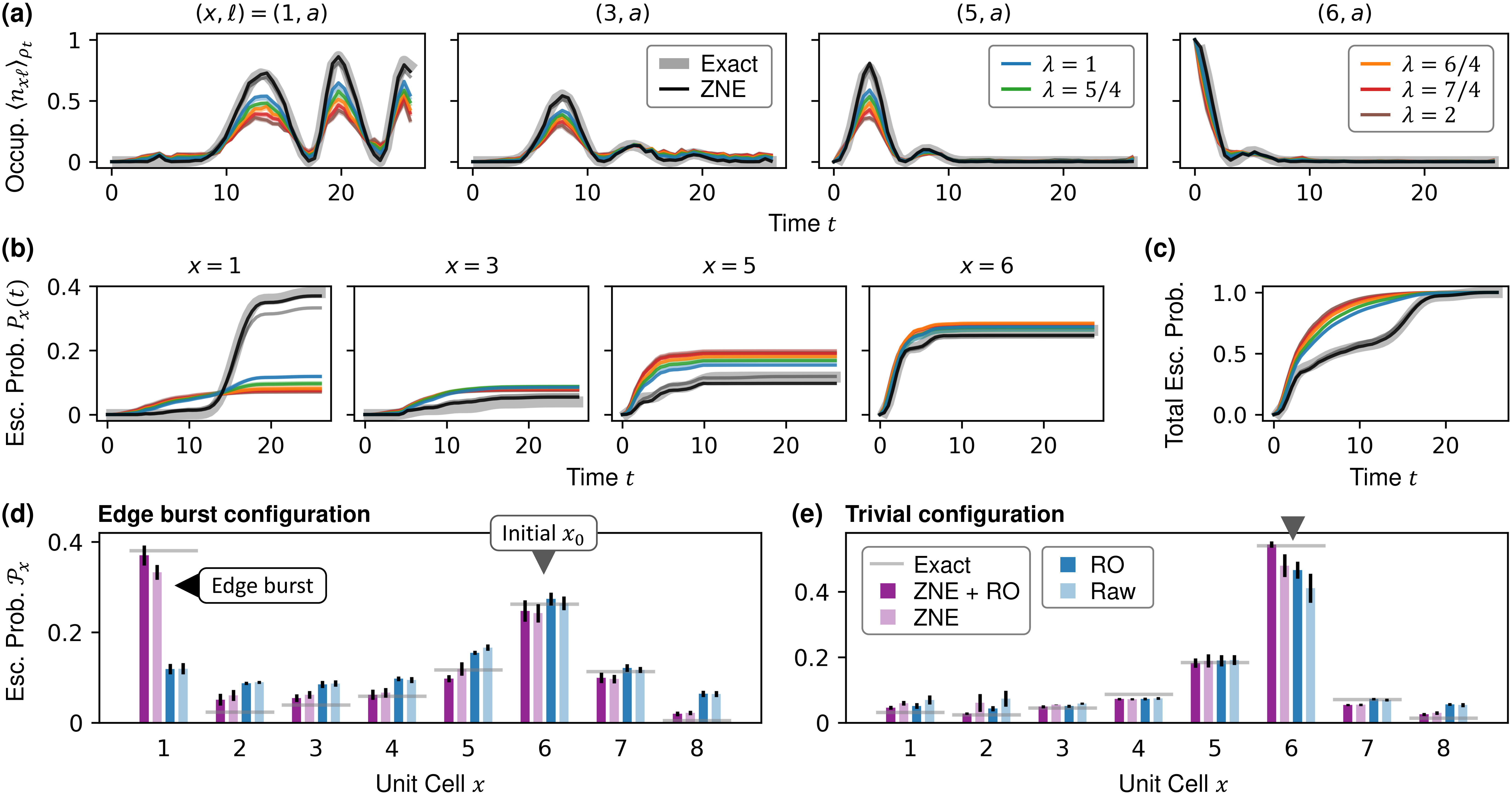}
    \phantomsubfloat{\label{fig:results-1p-small-system-a}}
    \phantomsubfloat{\label{fig:results-1p-small-system-b}}
    \phantomsubfloat{\label{fig:results-1p-small-system-c}}
    \phantomsubfloat{\label{fig:results-1p-small-system-d}}
    \phantomsubfloat{\label{fig:results-1p-small-system-e}}
    \vspace{-2\baselineskip}
    \caption{\textbf{Observing signatures of the non-Hermitian edge burst on a quantum processor.} \textbf{(a)} Site-resolved occupancy densities after normalized time-evolution on a $8$-unit cell quantum ladder, comparing data measured on quantum hardware at various noise amplification factors $\lambda$, after post-processing with zero-noise extrapolation (ZNE), and exact numerics. Solid thin lines are with readout error mitigation (RO) applied; translucent thin lines behind are without. To avoid clutter, only the $a$ sublattice of unit cells $x \in \{1, 3, 5, 6\}$ are shown. \textbf{(b)} Unit-cell-resolved escape probabilities $P_x(t)$ obtained from the data in (a). Time-integration of the occupancy densities first recovers proper wavefunction norm, and a second time-integration produces $P_x(t)$. \textbf{(c)} Total escape probability $P(t)$ summed over all unit cells, which approaches unity as time progresses. \textbf{(d)} Final unit-cell-resolved escape probabilities $\mathcal{P}_x$ in the long-time limit. Data obtained on quantum hardware with and without ZNE and RO, and from exact numerics, are shown. In a parameter regime supporting the edge burst, anomalously high escape probability on the $x = 1$ edge of the ladder is detected; whereas in the trivial regime escape probability is concentrated only near the initial location of the particle (at $x_0 = 6$). Error bars are standard deviations across $8$ experiment runs. See Supplementary Tables~\inertlink{S2} and \inertlink{S3} for Hamiltonian parameter values and superconducting quantum devices used.}
    \label{fig:results-1p-small-system}
\end{figure*}

Any attempt to realize time-evolution under a (arbitrary) non-Hermitian Hamiltonian $\mathcal{H}$ on a quantum platform runs invariably into a problem---that quantum states on a quantum device are by nature normalized, whereas $\mathcal{H}$ generates evolution that does not preserve normalization. Our approach is to realize the normalized evolution
\begin{equation}\begin{split}
    \rho_t
    = 
    \frac{V_t \rho_0 V_t^\dag}{\tr(V_t \rho_0 V_t^\dag)},
    \qquad 
    \rho_0 = \omega_0,
\end{split}\end{equation}
where $\smash{V_t = e^{-i \mathcal{H} t}}$ is the (non-unitary) time-evolution propagator over time $t$. Then the non-normalized physical state $\omega_t$ and $\rho_t$ are related by the rescaling $\omega_t = A_t^2 \rho_t$ for $A_t^2 = \smash{\tr(\omega_t)}$, and accordingly $\smash{\expval{O}_{\omega_t}} = A_t^2 \expval{O}_{\rho_t}$ for any observable $O$. Thus, in addition to implementing the normalized time-evolution, we develop a method to recover the normalization factor $A_t$ from measurements.

We employ a time-stepping algorithm to perform time-evolution from $\rho_0$ to $\rho_t$ (see \cref{fig:schematic/schematic-circuit-time-stepping}) that splits the evolution into $m$ steps, each approximately realizing time-evolution over an interval $\Delta t = t / m$ to an error $\order{\Delta t^2}$. Thus, the error accumulated over all steps scales as $\order{1 / m}$ and arbitrarily high simulation precision can be reached by increasing $m$. Writing $\mathcal{H} = \mathcal{H}_{\mathrm{H}} - i \mathcal{H}_{\mathrm{A}}$ where $\mathcal{H}_{\mathrm{H}}$ and $-i \mathcal{H}_{\mathrm{A}}$ are the Hermitian and anti-Hermitian parts of $\mathcal{H}$ respectively, each time step comprises the quantum maps $\mathcal{M}_{\mathrm{H}}$ and $\mathcal{M}_{\mathrm{A}}$ in sequence, designed to perform normalized time-evolution on arbitrary incident states by $\mathcal{H}_{\mathrm{H}}$ and $-i \mathcal{H}_{\mathrm{A}}$ respectively over time $\Delta t$ to error $\order{\Delta t^2}$.

The unitary channel $\mathcal{M}_{\mathrm{H}}$ can be implemented by standard methods. In the present work we use the first-order Trotter-Lie product formula~\cite{hatano2005finding}, which rewrites $\mathcal{H}_{\mathrm{H}}$ in the Pauli basis and performs time-evolution through a sequence of multi-qubit Pauli rotations (see \cref{fig:schematic/schematic-circuit-trotterization}). In general, higher-order product formulae or alternative unitary time-evolution circuit construction methods~\cite{ostmeyer2023optimised, ikeda2023trotter} can also be used. More details on the implementation of trotterized circuits on hardware are given in \methods. 

In contrast the map $\mathcal{M}_{\mathrm{A}}$, realizing action $\rho \mapsto \smash{V^{\mathrm{A}}_{\Delta t} \rho V^{\mathrm{A} \dag}_{\Delta t}} + \order{\Delta t^2}$ up to normalization for time-evolution propagator $\smash{V^{\mathrm{A}}_{\Delta t}} = \smash{e^{-\mathcal{H}_{\mathrm{A}} \Delta t}}$ and arbitrary incident states $\rho$, is non-unitary and is a source of difficulty. To perform $\mathcal{M}_{\mathrm{A}}$, we make use of the linear combination of unitaries (LCU) circuit primitive, which enables an effective action defined by sums of unitaries rather than products on a system register. In general, to effect an action $(\mu_1 U_1 + \cdots + \mu_d U_d)$ for coefficients $\{ \mu_k > 0 \}_k$ and unitaries $\{ U_k \}_k$ up to normalization, a register of $\ceil{\log_2 d}$ ancillary qubits and the ability to perform $U_k$ coherently controlled by the $\ket{k}$ state of the ancillary register suffice (see \cref{fig:schematic/schematic-circuit-lcu}). For our purpose, we approximate $\smash{V^{\mathrm{A}}_{\Delta t}}$ by a linear combination of forward and backward unitary time-evolution to error $\order{\Delta t^2}$. In fact, in general, for any $d \geq 2$, applicable mixtures of time-evolution operations can be efficiently found through power series expansions in $\Delta t$, and solutions of higher order error are available (see \methods).

To minimize the hardware resources required, we used $d = 2$ which necessitated only a single ancillary qubit. The pair of forward and backward time-evolution slices in the LCU are then defined by $\mu_+ = \mu_-$ and $U_\pm = \smash{e^{\pm i \mathcal{R} \Delta \tau}}$, for a rescaled time step $\Delta \tau = \smash{\sqrt{2 \Delta t}}$ and an auxiliary Hermitian Hamiltonian satisfying $\mathcal{R}^2 = \mathcal{H}_{\mathrm{A}}$. In the present edge burst setting, while choices of $\mathcal{R}^{\mathrm{eb}}$ satisfying $(\mathcal{R}^{\mathrm{eb}})^2 = \mathcal{H}_{\mathrm{A}}^{\mathrm{eb}}$ are readily found, further leveraging the structure of $\mathcal{H}_{\mathrm{A}}^{\mathrm{eb}}$ allowed refinements to $\mathcal{R}^{\mathrm{eb}}$ such that the LCU achieves the $\smash{V^{\mathrm{A}}_{\Delta t}}$ time-evolution exactly instead of to $\order{\Delta t^2}$ error (see \methods). We implemented the $U_\pm$ unitary time-evolution slices via first-order trotterization, with a low-overhead scheme in effecting the coherent controls by the ancillary qubit (see \methods).

The structure of the overall circuit we executed in experiments is shown in \cref{fig:schematic/schematic-circuit-experiment}. Through mid-circuit qubit reset, we re-initialized the ancillary qubit used for the LCU primitive between time steps---thus a single ancilla sufficed for the entire evolution from $\rho_0$ to $\rho_t$. Unlike other methods that can be adapted for non-Hermitian dynamical evolution~\cite{motta2020determining, sun2021quantum, mcardle2019variational, nishi2021implementation}, this protocol does not assume an ansatz for the time-evolved state, nor does it require iterative variational optimization or step-wise circuit construction based on intermediate measurements, which are often large in number. In \methods, we provide a generalized description of the time-stepping quantum simulation procedure outlined here. For cases where $\mathcal{R}_{\mathrm{A}}$ is difficult to obtain, we describe also a method based on a similar LCU expansion but requiring only access to terms in $\mathcal{H}_{\mathrm{A}}$. 

Upon completion of evolution to $\rho_t$, we simultaneously measured site-resolved occupancies $\smash{\expval{n_{x\ell}}_{\rho_t}}$ for all unit cells $x \in [N]$ and sublattices $\ell \in \{a, b\}$ on the quantum ladder. We then recovered the quantum state normalization factor $A_t$ via \cref{eq:results/model/wavefunction-norm} as the time integral
\begin{equation}\begin{split}
    A_t = \exp( -\gamma \int_0^t \sum_{x = 1}^N \expval{n_{x b}}_{\rho_\tau} \, \dd{\tau}),
    \label{eq:results/simulation/norm-time-integration}
\end{split}\end{equation}
which then enabled the rescaling of the measured $\smash{\expval{n_{x \ell}}_{\rho_t}}$ to $\smash{\expval{n_{x \ell}}_{\omega_t}}$. Lastly, a second time integration produces the escape probabilities $P_x(t)$ as in \cref{eq:results/model/escape-probs}. We run experiments to large times $t$ to access the long-time escape behavior $\mathcal{P}_x$ (see \methods).

We utilized IBM superconducting quantum processors in our experiments, which host up to $133$ transmon qubits connected in a heavy-hexagon topology with decoherence times $T_1, T_2$ up to ${\sim}\SI{200}{\micro\second}$. These devices support mid-circuit qubit readout used in our quantum circuits. Further details on the quantum hardware are provided in \methods, and device specifications such as gate and readout error rates are summarized in Supplementary Table~\inertlink{S4}. To address noise, which is non-negligible on present-day quantum hardware, we integrate several error suppression and mitigation methods such as dynamical decoupling, zero noise extrapolation with randomized gate twirling, and readout error mitigation (see \methods).

\subsection{Observing signatures of the non-Hermitian edge burst}
\label{sec:results/signatures}

To start, we realized the non-Hermitian quantum ladder $\mathcal{H}^{\mathrm{eb}}$ with $N = 8$ unit cells without interactions ($U_r = 0$ uniformly) and with open boundary conditions---natural on an open chain of qubits on the quantum processor---such that the ladder terminated in spatial boundaries past $x = 1$ and $x = 8$. The ladder is initialized with a particle on the $a$-sublattice of the $x = 6$ unit cell. We constructed our time-evolution circuits via trotterization as described earlier with standard transpilation onto hardware (see \methods), without employing more sophisticated circuit optimization techniques. 

We present a breakdown of experiment results in \cref{fig:results-1p-small-system}, illustrating the major steps in the measurement and data processing pipeline in addition to the final results. First, we show in \cref{fig:results-1p-small-system-a} site-resolved occupancy densities $\smash{\expval{n_{x \ell}}}_{\rho_t}$ on the normalized time-evolved quantum state $\rho_t$ as measured on the hardware. These occupancy densities were acquired at several noise amplification factors $\lambda \geq 1$, where $\lambda = 1$ corresponds to the circuits without modification and $\lambda > 1$ are on folded circuits of artificially increased gate count and depth (see \methods) but achieving, in the absence of noise, the same evolution. The increased circuit size amplifies noise suffered during time-evolution, arising from gate errors that are averaged by randomized gate twirling (see \methods) and decoherence of the qubits over time. The impact of the amplified noise is clear in the measured data: at higher $\lambda$, the amplitude of oscillations in $\smash{\expval{n_{x \ell}}}_{\rho_t}$ are suppressed and small-scale fluctuations become more prominent.

Employing a form of zero noise extrapolation (ZNE), we regressed the acquired data into the $\lambda = 0$ noiseless limit, taking into account physicality constraints such as particle number conservation and non-negativity of occupancy densities. As evident in \cref{fig:results-1p-small-system-a}, the $\smash{\expval{n_{x \ell}}}_{\rho_t}$ data after ZNE closely matched theoretical expectations calculated via exact diagonalization (ED) and presented a marked improvement over the $\lambda = 1$ data before ZNE. In addition to ZNE, we also employed readout error mitigation (RO), which approximately corrects bit-flip errors in measurement outcomes on the quantum processor (see \methods). Results without RO are inferior in accuracy and are shown in \cref{fig:results-1p-small-system-a} as translucent lines.

By performing time-integration on the $\smash{\expval{n_{x \ell}}}_{\rho_t}$ data, as described in \cref{eq:results/simulation/norm-time-integration}, we recovered the normalization factor of the quantum state and thereby rescaled $\smash{\expval{n_{x \ell}}}_{\rho_t}$ into occupancy densities $\smash{\expval{n_{x \ell}}}_{\omega_t}$ on the physical state $\omega_t$, the norm of which decreases over time (as $\mathcal{H}^{\mathrm{eb}}$ is lossy). A second time-integration on $\smash{\expval{n_{x \ell}}}_{\omega_t}$ yielded the escape probabilities $P_x(t)$ on unit cells $x$. We present these cell-resolved $P_x(t)$ in \cref{fig:results-1p-small-system-b}. The short-time increase of $P_x(t)$ at $x = 6$ expectedly arises from the initial localization of the particle on that unit cell, but as time progresses and the particle diffuses asymmetrically leftward under the NHSE, accumulation of $P_x(t)$ at cells of smaller $x$ occurs. We observe the impact of ZNE on the accuracy of the data, which brought the experiment $P_x(t)$ into close agreement with ED predictions. In \cref{fig:results-1p-small-system-c}, we report the total escape probability summed over all cells of the ladder, $P(t) = \smash{\sum_{x = 1}^N P_x(t)}$, which monotonically approached unity; the experiment was terminated when $P(t)$ reached sufficiently close to unity. 

Finally in \cref{fig:results-1p-small-system-d} we present cell-resolved final escape probabilities $\mathcal{P}_x$ along the ladder, showing raw experiment data, data with RO only, with ZNE only, and with both ZNE and RO applied. We had tuned $\mathcal{H}^{\mathrm{eb}}$ to be in a regime supporting the edge burst, here clearly manifesting as a prominent spike in $\mathcal{P}_x$ at the $x = 1$ boundary---larger, in fact, than at the $x_0 = 6$ initial localization of the particle. These results highlight the relevance of error mitigation: the signature of the edge burst is washed out without ZNE and RO, whereas with mitigation the escape probabilities closely agree with theory. Indeed, we emphasize the inherent sensitivity of the experimental measurements of $\mathcal{P}_x$: small inaccuracies in the $\smash{\expval{n_{x \ell}}}_{\rho_t}$ data, such as the suppressed oscillations in \cref{fig:results-1p-small-system-a}, can accumulate over the time-integration into large deviations in $P_x(t)$ and $\mathcal{P}_x$ at long times. This sensitivity places considerable demand on the robustness of the quantum simulation methodology and hardware, and underscores the difficulty of all experiments in our study.

In \cref{fig:results-1p-small-system-e} we report the counterpart of \cref{fig:results-1p-small-system-d} on $\mathcal{H}^{\mathrm{eb}}$ in the trivial regime, which does not exhibit the edge burst. The only peak in $\mathcal{P}_x$ occurs at the initial location of the particle, reflecting the decay of the particle in-situ with no physically significant dynamics over time.

We describe additional experiments on the same $N = 8$ ladder but with initial particle localization at $x_0 = 4$, and on a smaller $N = 4$ ladder, in Supplementary Note~\inertlink{4A}. The conclusions from those results are qualitatively identical to those drawn here, and the real-space signature of the edge burst in $\mathcal{P}_x$ were likewise clearly observed.

\subsection{Edge burst on larger system sizes}
\label{sec:results/large-system}

\begin{figure*}[!t]
    \centering
    \includegraphics[width = \linewidth]{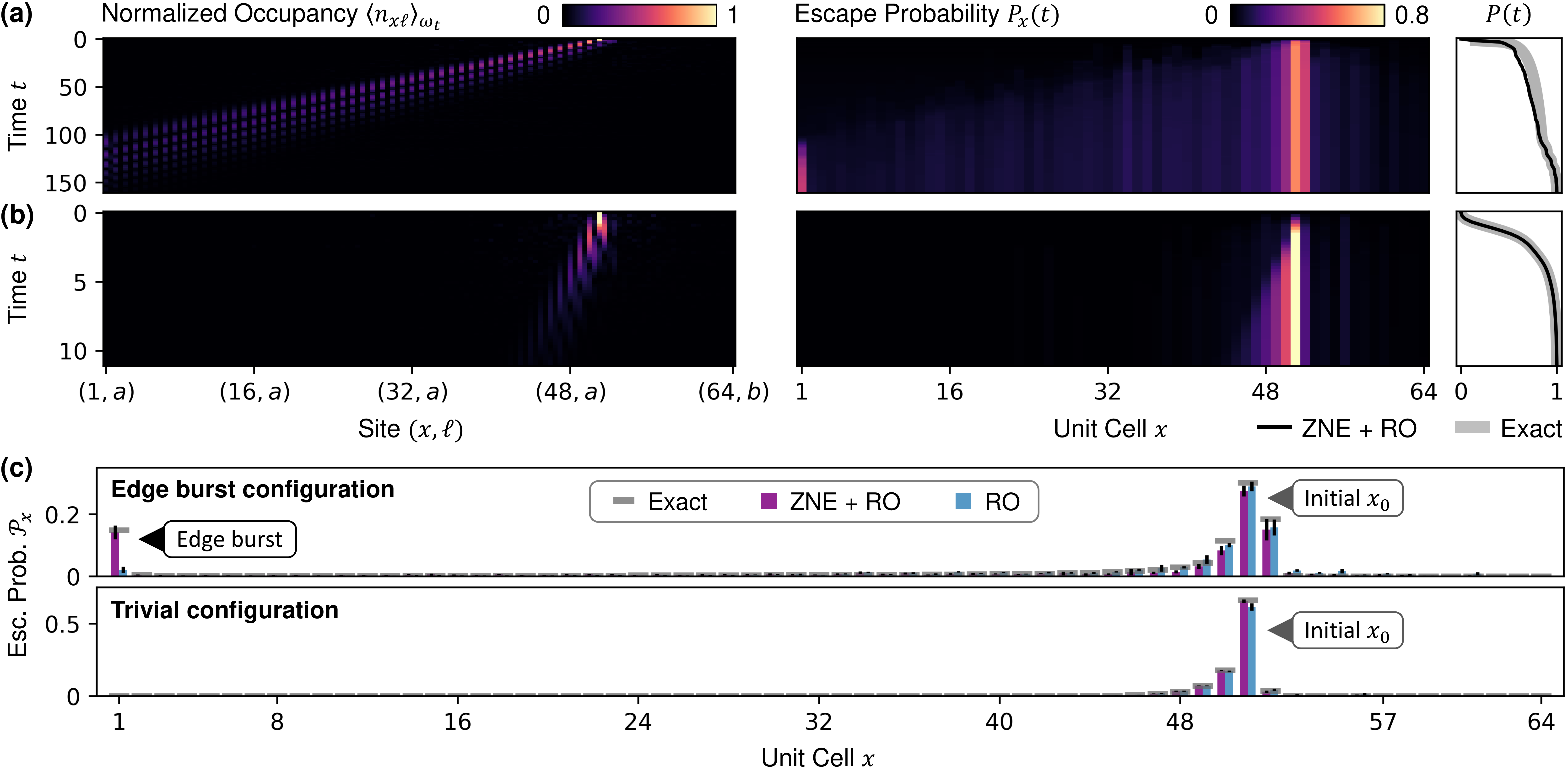}
    \phantomsubfloat{\label{fig:results-1p-large-system-a}}
    \phantomsubfloat{\label{fig:results-1p-large-system-b}}
    \phantomsubfloat{\label{fig:results-1p-large-system-c}}
    \vspace{-2\baselineskip}
    \caption{\textbf{Physical demonstration of edge burst at large system size.} \textbf{(a)} Site-resolved occupancy densities and unit-cell-resolved escape probabilities $P_x(t)$ over time on a $64$-unit cell quantum ladder in the edge burst regime measured on quantum hardware, with zero-noise extrapolation (ZNE) and readout error mitigation (RO) applied. The particle was initially localized at $x_0 = 51$. Proper quantum state normalization was recovered via time-integration of measured occupancy densities and applied to both panels. The total escape probability $P(t)$ approaches unity as time progresses. \textbf{(b)} Same as (a) but in the trivial regime. \textbf{(c)} Final unit-cell-resolved escape probabilities $\mathcal{P}_x$ in the long-time limit, comparing results in the edge burst regime and in the trivial regime. Data obtained on quantum hardware with ZNE and RO applied, RO only, and exact numerics are shown. Error bars are standard deviations across $16$ experiment runs. See Supplementary Tables~\inertlink{S2} and \inertlink{S3} for Hamiltonian parameter values and superconducting quantum devices used.}
    \label{fig:results-1p-large-system}
\end{figure*}

We next investigated larger system sizes. Aside from demonstrating the versatility of our quantum simulation approach, experiments on larger ladders also minimize finite-size effects and produce clearer evidence of the non-Hermitian edge burst. Here we realize an $N = 64$ unit cell ($128$-site) ladder, which is $8$ times the system size examined in the previous section.

At such large system sizes, trotterization and standard circuit transpilation produce circuits far too deep for present NISQ hardware to feasibly accommodate---naïve execution of these circuits would result in near-complete decoherence of the quantum state and vanishing signal-to-noise ratios when measuring observables. To overcome this present limitation, we employed an additional tensor-network aided circuit recompilation technique~\cite{koh2024realization, koh2022simulation, koh2022stabilizing, sun2021quantum} for circuit compression, which replaces components of the circuits with approximate lower-depth parametrized ansatzes that are variationally optimized (see \methods). In this process, we exploited symmetries of $\mathcal{H}^{\mathrm{eb}}$, such as number conservation, to enhance circuit construction performance and quality.

In the left panel of \cref{fig:results-1p-large-system-a}, we report site-resolved occupancy densities $\smash{\expval{n_{x \ell}}}_{\omega_t}$ on the $N = 64$ ladder in the edge burst regime, as obtained on hardware with ZNE and RO error mitigation applied. Underlying this data were $\smash{\expval{n_{x \ell}}}_{\rho_t}$ measurements that had been time-integrated. As $\mathcal{H}^{\mathrm{eb}}$ is lossy, the occupancy densities $\smash{\expval{n_{x \ell}}}_{\omega_t}$ decay with time, reflecting particle escape from the ladder. In the right panel we report the cell-resolved escape probabilities $P_x(t)$ time-integrated from $\smash{\expval{n_{x \ell}}}_{\omega_t}$. 

For comparison, we show in \cref{fig:results-1p-large-system-b} the counterpart to \cref{fig:results-1p-large-system-a} for $\mathcal{H}^{\mathrm{eb}}$ in the trivial regime. In both set-ups, asymmetrical leftwards drift of the particle along the ladder driven by the NHSE is clearly observed; but in the edge burst regime the decay of the particle is significantly slower, allowing the particle to reach the $x = 1$ boundary with non-negligible amplitude (\ie~survival probability). Both experiments were terminated when the total escape probability $P(t)$ reached close to unity.

Finally, we present the final escape probabilities $\mathcal{P}_x$ in \cref{fig:results-1p-large-system-c} for both regimes. As before, the edge burst manifests as a prominent spike in escape probability at the $x = 1$ boundary, which is absent in the trivial regime. This demanding setting at large system size reinforces the relevance of a robust quantum simulation methodology integrated with effective error mitigation: the experimental $\mathcal{P}_x$ data closely agrees with theory with mitigation fully employed.

\subsection{Spectral information of the edge burst}
\label{sec:results/spectral}

\begin{figure}[!t]
    \centering
    \phantomsubfloat{\label{fig:results-1p-spectral-circuit}}
    \phantomsubfloat{\label{fig:results-1p-spectral-extremal-Im-E}}
    \phantomsubfloat{\label{fig:results-1p-spectral-pbc-spectrum}}
    \includegraphics[width = \linewidth]{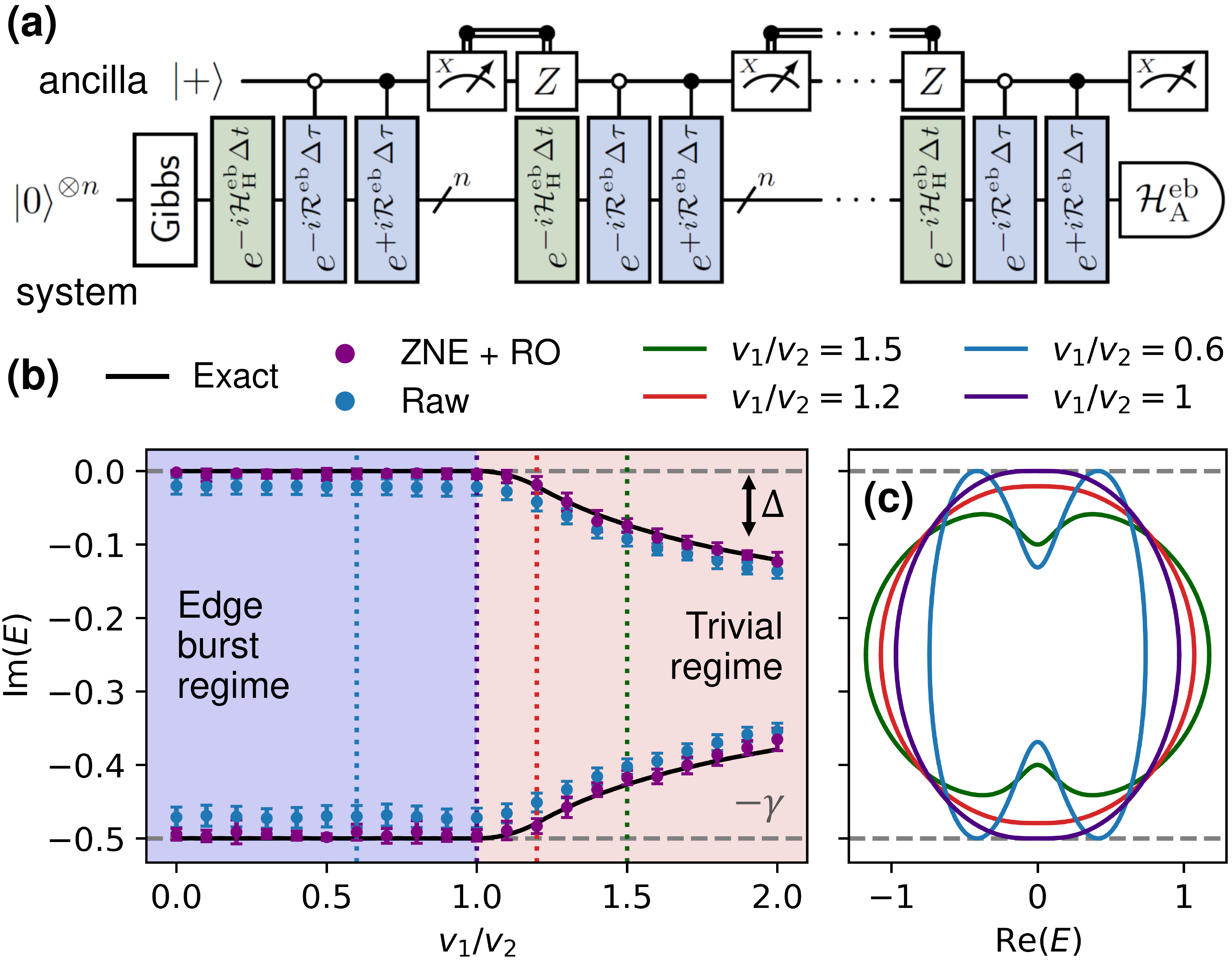}
    \caption{\textbf{Probing spectral properties of the edge burst system.} \textbf{(a)} Schematic of the circuits used to measure extremal imaginary energy. The system is prepared in a Gibbs state and time-evolved under the non-Hermitian $\mathcal{H}^{\mathrm{eb}}$ for long times. The expectation value of the anti-Hermitian component of the Hamiltonian $-i \mathcal{H}_{\mathrm{A}}^{\mathrm{eb}}$ is measured on the time-evolved state. \textbf{(b)} Maximum and minimum imaginary energy versus hopping amplitudes $v_1 / v_2$, showing hardware data measured on an $N = 16$ quantum ladder with and without ZNE and RO, and exact numerics. The $v_1 / v_2 \leq 1$ regime, wherein the dissipative gap $\Delta$ closes, supports the edge burst; the $v_1 > v_2$ regime is trivial. Error bars are standard deviations across $25$ experiment runs. See Supplementary Tables~\inertlink{S2} and \inertlink{S3} for Hamiltonian parameter values and superconducting quantum devices used. \textbf{(c)} Complex energy spectra of $\mathcal{H}^{\mathrm{eb}}$ along indicated values of $v_1 / v_2$ from numerics. The maxima and minima of these spectra are traced in (b). The horizontal and vertical reflection symmetries of the spectra are protected by symmetries of the Hamiltonian.}
    \label{fig:results-1p-spectral}
\end{figure}

Time-evolution on the non-Hermitian quantum ladder Hamiltonian $\mathcal{H}^{\mathrm{eb}}$ comprises conceptually of two intermixed components---real time-evolution on the Hermitian part of the Hamiltonian $\mathcal{H}^{\mathrm{eb}}_{\mathrm{H}}$ and imaginary time-evolution on $\mathcal{H}^{\mathrm{eb}}_{\mathrm{A}}$. Unlike real-time evolution, imaginary time-evolution on a Hermitian Hamiltonian is not energy-conserving and, in fact, yields quantum states of extremal energy at long times. Thus, by evolving to long times under $\mathcal{H}^{\mathrm{eb}}$, we can purify a starting state into an eigenstate of extremal imaginary eigenenergy.

An illustration of this type of quantum circuits we executed is shown in \cref{fig:results-1p-spectral-circuit}. To ensure nonzero overlap with the extremal eigenstates, we initiated time-evolution with the infinite-temperature Gibbs state (\ie~maximally mixed state). This is prepared through a completely depolarizing channel implemented in constant depth through a single round of mid-circuit measurements whose outcomes are discarded (see \methods). The time-evolution quantum algorithm for non-Hermitian Hamiltonians, as described earlier (drawn in \cref{fig:schematic/schematic-circuit-time-stepping}), is invoked following state preparation. After evolution for a sufficiently long time, we measured the imaginary energy of the quantum state through Hamiltonian averaging on $\mathcal{H}^{\mathrm{eb}}_{\mathrm{A}}$ (see \methods).

We report the extremal imaginary eigenenergies $\Im E$ measured on hardware as a function of hopping amplitudes $v_1 / v_2$ in \cref{fig:results-1p-spectral-extremal-Im-E}. Of particular relevance is the largest imaginary eigenenergy $\max \Im E$, which is zero for $v_1 \leq v_2$ but strictly negative for $v_1 > v_2$. That is, the imaginary gap $\Delta = -\max \Im E$, also known as the dissipative gap, of $\mathcal{H}^{\mathrm{eb}}$ closes for $v_1 \leq v_2$. In \cref{fig:results-1p-spectral-pbc-spectrum} we illustrate the complex energy spectrum of $\mathcal{H}^{\mathrm{eb}}$ at various $v_1 / v_2$, which makes clear the recession of the spectrum into the negative imaginary half-plane as $v_1$ exceeds $v_2$. We remark that the horizontal and vertical reflection symmetries of the complex spectra, about $\Re E = 0$ and $\Im E = -\gamma / 2$, are consequent of chiral and time-reversal symmetries of the quantum ladder model (see Supplementary Note~\inertlink{1A}). 

Prior theoretical analysis~\cite{xue2022non} indicated that the closing of the dissipative gap $\Delta$ is a necessary condition for the edge burst to occur in addition to the presence of the NHSE. Intuitively, the zero imaginary eigenenergy modes suffer no decay during time-evolution and are long-lived, thus enabling an initial wavepacket driven by the NHSE to reach the boundary of the system. The particle is then trapped against the boundary by the NHSE and decays, giving rise to the edge burst. In contrast, upon opening of the dissipative gap, any wavepacket suffers exponential loss during propagation and cannot reach the boundary before decay. This qualitative difference was clearly observed, for example, in our experiment results in \cref{fig:results-1p-large-system}. Indeed, in all our experiments the edge burst regime---where edge burst signatures are present---occur when $v_1 \leq v_2$, and the trivial regimes occur when $v_1 > v_2$.

\subsection{Spatially extended and ordered edge bursts with multiple interacting particles}
\label{sec:results/interacting-spatial}

\begin{figure*}[!t]
    \centering
    \includegraphics[width = \linewidth]{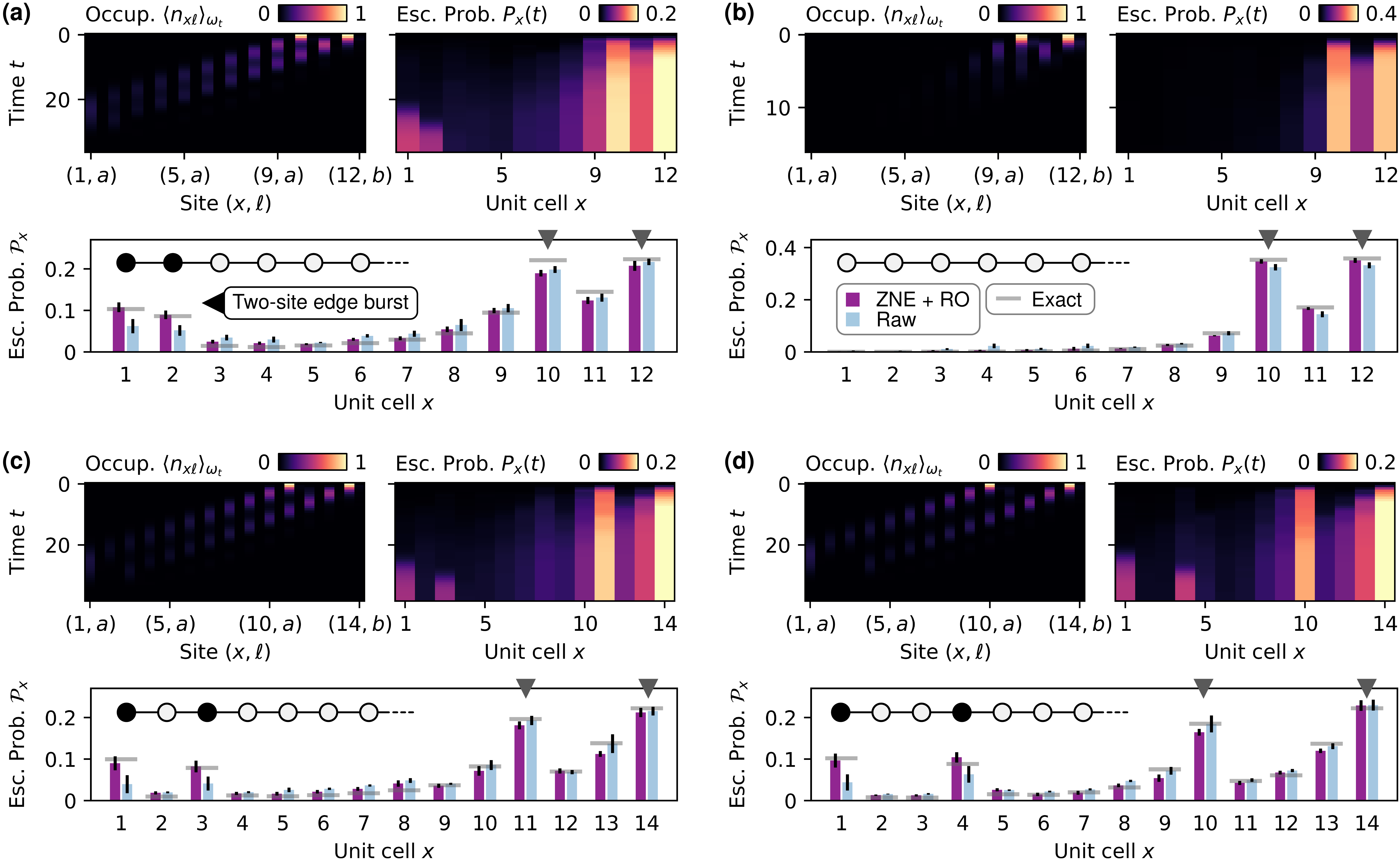}
    \phantomsubfloat{\label{fig:results-2p-R1-burst}}
    \phantomsubfloat{\label{fig:results-2p-R1-nobur}}
    \phantomsubfloat{\label{fig:results-2p-R3-burst}}
    \phantomsubfloat{\label{fig:results-2p-R5-burst}}
    \vspace{-1.6\baselineskip}
    \caption{\textbf{Spatially extended edge bursts with multiple interacting particles.} \textbf{(a)} Site-resolved occupancy densities and unit-cell-resolved escape probabilities $P_x(t)$ measured on a $12$-unit cell quantum ladder in the edge burst regime hosting two interacting particles. Final escape probabilities $\mathcal{P}_x$ obtained on hardware with and without zero-noise extrapolation (ZNE) and readout error mitigation (RO), and exact numerics, are shown. An edge burst spatially extended across two unit cells is observed, highlighted as black shading in the cartoon inset. \textbf{(b)} Same as (a) but in the trivial regime, which does not give rise to an edge burst. \textbf{(c)--(d)} By increasing the range of interactions on the quantum ladder, spatially extended edge bursts separated by one or more unit cells arise. Gray arrows denote the initial localization of the particles. Error bars are standard deviations across $10$ experiment runs. See Supplementary Tables~\inertlink{S2} and \inertlink{S3} for Hamiltonian parameter values and superconducting quantum devices used.}
    \label{fig:results-2p}
\end{figure*}

\begin{figure*}
    \centering
    \includegraphics[width = \linewidth]{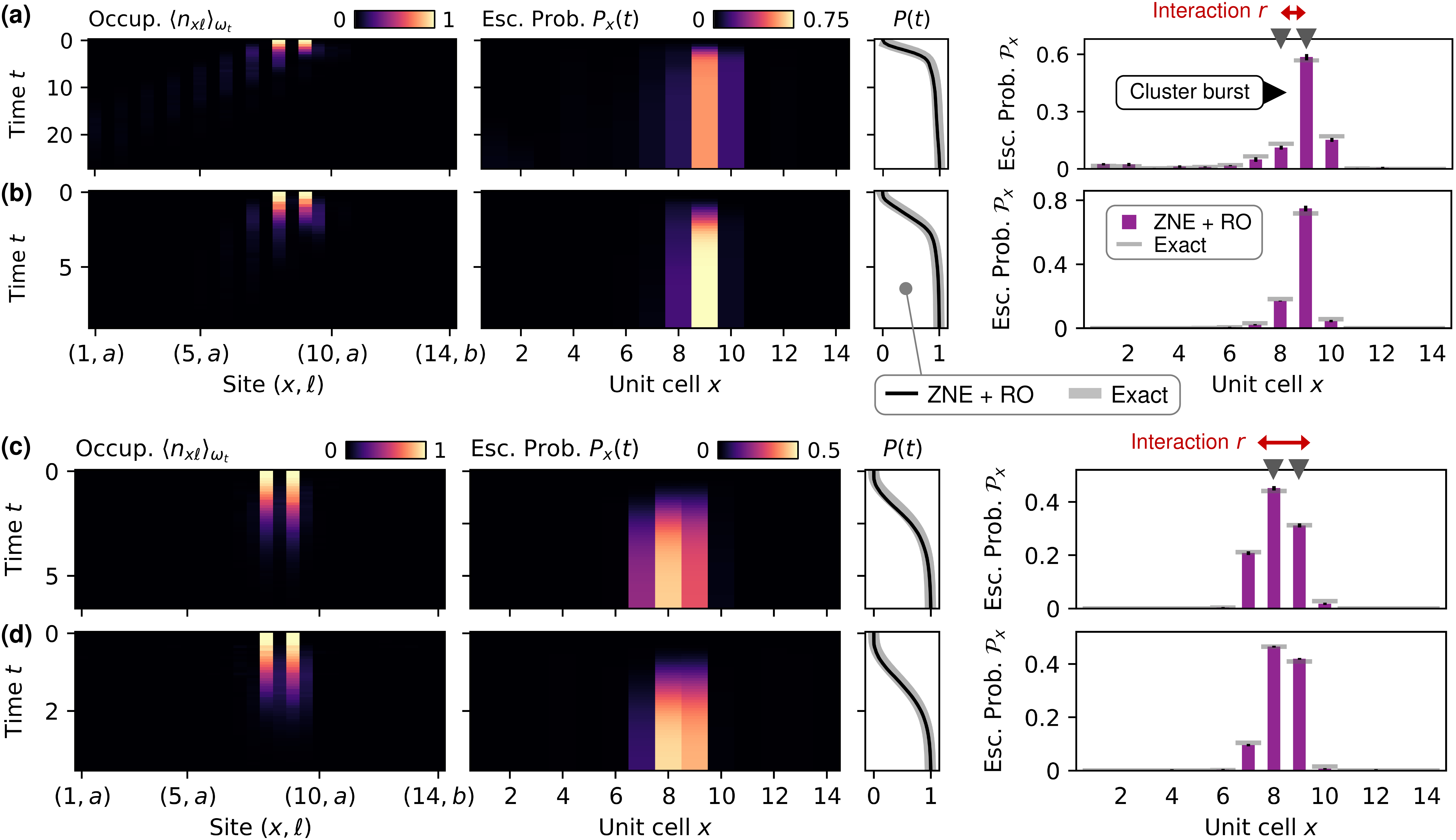}
    \phantomsubfloat{\label{fig:results-cluster-R1-OR-burst}}
    \phantomsubfloat{\label{fig:results-cluster-R1-OR-nobur}}
    \phantomsubfloat{\label{fig:results-cluster-R5-IR-burst}}
    \phantomsubfloat{\label{fig:results-cluster-R5-IR-nobur}}
    \vspace{-1.6\baselineskip}
    \caption{\textbf{Cluster bursts away from boundaries induced by interactions.} \textbf{(a)} Site-resolved occupancy densities, unit-cell-resolved escape probabilities $P_x(t)$, and total escape probability $P(t)$ summed over all unit cells, measured on a $14$-unit cell quantum ladder in the edge burst regime hosting two interacting particles. Final escape probabilities $\mathcal{P}_x$ obtained on hardware with zero-noise extrapolation (ZNE) and readout error mitigation (RO), and exact numerics, are shown. Initial localizations of the two particles (gray triangles) are barely outside the range of interactions (red arrow). \textbf{(b)} Same as (a) but with the quantum ladder in the trivial regime. \textbf{(c)--(d)} Measurements on the same quantum ladder as (a)--(b) but with the initial localizations of the particles (gray triangles) inside the range of interactions (red arrow), with the ladder in (c) the edge burst regime and (d) the trivial regime. In all cases, a cluster burst is observed in the vicinity of the initial locations of the particles, far from boundaries. Error bars are standard deviations across $10$ experiment runs. See Supplementary Tables~\inertlink{S2} and \inertlink{S3} for Hamiltonian parameter values and superconducting quantum devices used.}
    \label{fig:results-cluster}
\end{figure*}

We now turn to our key set of results: novel edge burst phenomena observed with multiple interacting particles. Systems hosting strongly interacting quantum particles are difficult to realize on classical or effectively single-body quantum simulators, such as electrical (\ie~topolectrical) circuits~\cite{Helbig2020, Lee2018, Zou2021, zhang2023} or waveguide photonic systems~\cite{Alaeian2014, Li2023_ep, nasari2023non}, that have thus far been used to study non-Hermitian physics---let alone achieving tunability of interaction strengths, ranges and types. In this respect digital quantum simulation on quantum hardware, making use of the full many-body Hilbert space hosted by the device and programmable quantum operations in that space, presents a clear versatility advantage. The quantum simulation approach we developed accommodates arbitrary interactions in the Hamiltonian (see \methods), and here we exploit this versatility to explore non-Hermitian edge burst phenomenology enriched by interactions.

To start, we examined two particles (hardcore bosons) on an $N = 12$ ladder in the edge burst regime with $U_1 > 0$ interactions switched on---that is, repulsive density-density interactions acting between sites on the ladder a range $r = 1$ apart (see \cref{fig:schematic/schematic-model-interactions} for illustration). The energy scale of the interaction was comparable to those of the non-interacting parts of the Hamiltonian (hoppings $v_1, v_2$ and loss $\gamma$). We present experiment results in \cref{fig:results-2p-R1-burst}, showing the measured site-resolved occupancy densities $\smash{\expval{n_{x \ell}}}_{\omega_t}$, recovered cell-resolved escape probabilities $P_x(t)$, and final escape probabilities $\mathcal{P}_x$. As in our prior experiment on the vanilla edge burst at large system size, we employed circuit recompilation to compress circuit depth and utilized ZNE and RO error mitigation to address hardware noise. 

Here, interestingly, from $\mathcal{P}_x$ we observe the manifestation of a spatially extended version of the edge burst, spanning two unit cells ($x \in \{1, 2\}$) instead of the typical single unit cell in vanilla edge bursts. The mechanism underlying this spatial extension stems from a Pauli exclusion-like effect arising from the interactions. At short times both particles expectedly diffuse to the left under the NHSE, but as the first impacts on the $x = 1$ boundary, the hardcore nature of the bosons and the $U_1$ interactions inhibit the second from also occupying the first unit cell---effectively, the second boson encounters a virtual boundary at $x = 2$. This is akin to the formation of the so-called Fermi skin in fermionic non-Hermitian systems~\cite{lee2021many, shen2022non}, wherein fermions are pushed up against a boundary by the NHSE and thereby exhibit a Fermi surface in real-space. Here, the second boson is trapped at the $x = 2$ unit cell and decays, giving rise to a spatially extended edge burst occurring over two unit cells. In the trivial regime the edge burst does not manifest, as shown in \cref{fig:results-2p-R1-nobur}.

Moreover, we observed that by tuning the range of interactions, spatially ``ordered'' manifestations of the edge burst can arise. In \cref{fig:results-2p-R3-burst} we present experiment results on an $N = 14$ ladder hosting two bosons with $U_r > 0$ switched on for $r \leq 3$, and in \cref{fig:results-2p-R5-burst} we present results with $U_r > 0$ for $r \leq 5$. In the former we witness a spatially extended edge burst occupying alternate unit cells, and in the latter the edge burst is separated by two unit cells of small escape probability. The cartoon insets in \cref{fig:results-2p-R3-burst,fig:results-2p-R5-burst} highlight these patterns. In analogy to spatial ordering induced by interactions in conventional (Hermitian) spin and atomic systems, we refer to these as $\mathbb{Z}_2$ and $\mathbb{Z}_3$ orders respectively. The mechanism underlying their formation is similar to the $U_1$ setting, but here the longer-ranged interactions induce virtual boundaries located farther from the first boson at $x = 1$. 

The same phenomenon occurs when $p > 2$ particles are present, which gives rise to spatially extended edge bursts occurring over $p$ unit cells. Generally, the presence of $U_r > 0$ interactions for $r \leq 2 d - 1$ can produce $\mathbb{Z}_d$ edge burst orderings; with $p$ bosons, the same $\mathbb{Z}_d$ pattern occurs $p$ times starting from the left boundary of the ladder (the drive direction of the NHSE). In Supplementary Note~\inertlink{4C}, we describe additional experiments conducted with three bosons yielding a three-unit-cell spatially extended edge burst. We also observed qualitatively similar $\mathbb{Z}_2$ and $\mathbb{Z}_3$ orderings under longer-range interactions on the three-boson setup. Likewise we verified that the edge bursts do not arise in the trivial regime.

In the settings examined here (in \cref{fig:results-2p}), the initial conditions were such that the bosons do not significantly infringe into the spatial ranges of $U_r$ interactions under the dynamics of the system---\ie~they do not approach too closely to one another---until they impact on the left boundary. This allows the discussed physical mechanism to carry through unimpeded. As the leftward drift induced by the NHSE is approximately uniform, this condition is simple to achieve and boils down to the initial localization of the bosons. Indeed, in \cref{fig:results-2p-R1-burst,fig:results-2p-R3-burst,fig:results-2p-R5-burst}, with $U_r > 0$ for $r \leq 2 d - 1$, the bosons were initially separated by at least $2d + 1$ sites on the ladder or equivalently ${\sim}d$ unit cells. When this condition is not satisfied, a different type of interaction-driven edge burst phenomenology can occur, as we describe next.

\subsection{Interaction-induced cluster bursts}
\label{sec:results/interacting-cluster}

Finally, we demonstrate that with appropriately configured interactions, anomalously high escape probabilities can be induced at chosen locations on the quantum ladder away from boundaries, in fact regardless of whether the Hamiltonian is in the (canonical) edge burst or trivial regimes. We term this phenomenon as cluster bursts. Unlike the canonical edge burst, cluster bursts need not occur at the edges of the system and arise intrinsically from interactions; there is no analogous single-body non-interacting counterpart known at present.

We show in \cref{fig:results-cluster-R1-OR-burst} experiment results on an $N = 14$ ladder in the canonical edge burst regime ($v_1 \leq v_2$) hosting two bosons. The bosons were initially localized on the $a$-sublattices of unit cells $x_0 = 8, 9$, and we enable density-density interaction $U_1$ of energy scale comparable to the non-interacting parts of the Hamiltonian ($v_1, v_2, \gamma$). Thus, the initial conditions are such that the particle separation condition discussed in the previous section is not clearly satisfied---in particular a single hop of either boson toward the other, via either $v_1$ or $v_2$ channels, places them within the range of interactions and causes a significant energy change. Strikingly we observe in experimentally measured $\mathcal{P}_x$ that both bosons decay essentially \textit{in-situ}, with an overwhelming proportion of escape probability concentrated precisely at $x = 8, 9$. The same cluster burst arises when the Hamiltonian is tuned into the trivial regime, as we show in \cref{fig:results-cluster-R1-OR-nobur}, with qualitatively similar features in $\mathcal{P}_x$.

This same phenomenon can also arise when the initial localizations of the bosons are inside the range of interactions, such that outward hopping puts them outside their interaction range and causes an energy change. In this case, the bosons decay and a cluster burst emerges within the spatial confines of the interaction range. We show an example on the same $N = 14$ ladder in \cref{fig:results-cluster-R5-IR-burst,fig:results-cluster-R5-IR-nobur} tuned into the edge burst and trivial regimes respectively. As before the bosons were initialized on unit cells $x_0 = 8, 9$, but here interactions $U_r > 0$ were switched on for $r \leq 3$. In both edge burst and trivial regimes, we observe clearly a cluster burst in the vicinity of the initial boson locations, far from boundaries.

The mechanism underlying these cluster bursts stem from energetically (partially) forbidden transitions caused by the interactions. We remind that energy is conserved in time translation-symmetric closed quantum systems---\ie~throughout time-evolution under a Hermitian Hamiltonian $\mathcal{H}$, the energy $\tr(\omega_t \mathcal{H})$ of a quantum state $\omega_t$ is fixed. It is then a well-established effect that strong interactions can cause energy differences between states connected by the Hamiltonian so large as to essentially forbid transitions out of or from a quantum state, as there is no available linear combination of accessible states after the transition that maintains the conserved energy. Indeed, this mechanism underlies a doublon decay phenomenon observed on Hubbard models~\cite{yin2023prethermalization, sensarma2010lifetime}, where pairs of excitations in proximity exhibit exponentially long lifetimes, and is responsible also for the well-studied Stark many-body localization~\cite{schulz2019stark, morong2021observation}. 

A similar understanding holds in a non-Hermitian context, with the modification that energy is not exactly conserved but there exists nonetheless a speed limit on how fast energy can change in the system, imposed by the Hamiltonian (see Supplementary Note~\inertlink{1B}). On the quantum ladder, this slow-down in dynamics freezes the bosons long enough that they decay largely in their initial locations, generating the cluster burst observed. In the scenarios demonstrated in \cref{fig:results-cluster}, for example, two simultaneous hoppings of the bosons must occur for them to move along the NHSE drive direction without changing their interaction energy, which is a second-order process and happens with low amplitude (probability). In the unlikely event that these higher-order hoppings occur and the bosons reach the boundary of the system, spatially extended edge bursts of small amplitude can additionally manifest, as can also be observed in \cref{fig:results-cluster-R1-OR-burst}.

\section{Discussion}
\label{sec:conclusion}

While the experimental frontier of non-Hermitian condensed-matter physics has to-date enjoyed a spectacular period of progress utilizing broad palettes of custom-designed analog optical, metamaterial, and classical simulator platforms, the direct realization of important classes of non-Hermitian systems on near-term (\ie~NISQ) quantum platforms has remained very limited. The advantages offered by digital quantum simulation are clear and tantalizing: the natural ability to access the many-body Hilbert space of arbitrary Hamiltonians, and in association, liberal versatility in tuning interactions of different ranges, strengths and types. 

But leveraging quantum simulators to probe non-Hermitian systems is challenging, as quantum operations on quantum devices are unitary and thus sophisticated methods must be developed to achieve the non-unitary evolution associated with non-Hermitian Hamiltonians. Moreover, the overhead incurred in doing so invariably runs up against qubit error and lifetime limitations. Nevertheless, recent studies of various condensed-matter phenomena on digital quantum computers, ranging from discrete time crystals~\cite{mi2022time, frey2022realization, chen2023robust} to topological phases~\cite{azses2020identification, mei2020digital, koh2022stabilizing, koh2024realization, koh2022simulation, mi2022time, satzinger2021realizing, google2023non, iqbal2024topological}, illustrate promising capabilities despite hardware constraints. It is therefore timely to investigate the prospects of studying non-Hermitian quantum physics on quantum hardware.

Here, we probed the recently discovered phenomenon of non-Hermitian edge bursts on transmon-based superconducting quantum processors, and observed at high fidelity the signatures of the edge burst on up to $64$ unit cells of a lossy quantum ladder model. Furthermore, by incorporating sequences of density-density interactions, we unveiled the possibility of engineering spatially extended and ordered variants of the edge burst, as well as bursts occurring away from boundaries, which we dubbed cluster bursts. This marks the first time the non-Hermitian edge burst has been realized on an intrinsically quantum platform, as well as the first experimental study of interacting variants of the edge burst, complementing a prior experimental effort on the canonical non-interacting version of the edge burst~\cite{xiao2024observation,zhu2024observation}. To enable this advance, we developed a general methodology for efficient non-Hermitian Hamiltonian simulation on digital quantum processors, leveraging a linear combination of unitaries circuit construction technique with ancillary qubit re-use, which is applicable to generic non-Hermitian models far beyond the scope of the edge burst.

Our work opens the door for future experimental investigation of quantum non-Hermitian condensed-matter. As quantum hardware continues to advance, we anticipate that our methods will enable the direct study of more sophisticated systems exhibiting rich intertwined physics, such as critical versions of non-Hermitian pumping~\cite{Li2020critical} and the interplay between non-Hermiticity and entanglement phase transitions~\cite{kawabata2023entanglement}.

\section{Acknowledgements}

J.M.K.~thanks Jayne Thompson, Jun Ye, and Jian Feng Kong of the Quantum Innovation Center (Q.Inc) and Institute of High Performance Computing (IHPC), Agency for Science, Technology and Research (A*STAR), Tianqi Chen of the National University of Singapore, and Mincheol Park of Harvard University, for helpful discussions. The authors acknowledge the use of IBM Quantum services for this work. The views expressed are those of the authors, and do not reflect the official policy or position of IBM or the IBM Quantum team. J.M.K.~and T.T.~are grateful for support from the A*STAR Graduate Academy. D.E.K.\ is supported by the National Research Foundation, Singapore, and the Agency for Science, Technology and Research (A*STAR), Singapore, under its Quantum Engineering Programme (NRF2021-QEP2-02-P03); A*STAR C230917003; and A*STAR under the Central Research Fund (CRF) Award for Use-Inspired Basic Research (UIBR) and the Quantum Innovation Centre (Q.InC) Strategic Research and Translational Thrust. We acknowledge support from the  Ministry of Education, Singapore Tier-II grant (MOE award number: MOE-T2EP50222-0003).

\bibliography{ref-qc,ref-topo}

\clearpage
\pagebreak

\small

\section*{Methods}
\label{sec:methods}

\textbf{Quantum hardware.}
\label{sec:methods/hardware}
We used IBM transmon-based superconducting quantum devices in our experiments. These included 27-qubit devices \textit{ibm\_hanoi} and \textit{ibm\_mumbai} hosting the Falcon processor, 127-qubit devices \textit{ibm\_sherbrooke}, \textit{ibm\_osaka}, \textit{ibm\_kyoto} and \textit{ibm\_nazca} hosting the Eagle processor, and a 133-qubit device \textit{ibm\_torino} hosting the Heron processor. The basis gate sets of all devices comprise 1-qubit gates $\smash{\{\mathrm{X}, \sqrt{\mathrm{X}}, \mathrm{RZ}\}}$, with $\mathrm{RZ}$ implemented virtually via framechanges~\cite{mckay2017efficient}, and 2-qubit gate CX for Falcon processors, echoed cross resonance (ECR) for Eagle processors~\cite{sundaresan2020reducing}, and CZ for the Heron processor. The CX, ECR and CZ gates are equivalent up to 1-qubit rotations. We construct all experiment circuits using CX gates, including the application of protocols for error suppression and mitigation such as Pauli twirling and randomized gate folding for zero-noise extrapolation (see \hyperref[sec:methods/pauli-twirling]{below}), and transpile to the native 2-qubit gate of each device before execution. Typical performance metrics of the devices, such as relaxation $T_1$ and dephasing $T_2$ times, gate and readout error rates, and gate times are provided in Supplementary Table~\inertlink{S4}.

\textbf{Model.} 
\label{sec:methods/model}
We examined the bosonic quantum ladder Hamiltonian $\mathcal{H}^{\mathrm{eb}} = \mathcal{H}^{\mathrm{eb}}_0 + \mathcal{H}^{\mathrm{eb}}_{\mathrm{int}}$, where $\mathcal{H}^{\mathrm{eb}}_0$ and $\mathcal{H}^{\mathrm{eb}}_{\mathrm{int}}$ comprise non-interacting and interacting terms respectively, given by
\begin{equation}\begin{split}
    \mathcal{H}^{\mathrm{eb}}_0
    = \sum_{x = 1}^N \bigg\{ 
        & -i \gamma c^\dag_{x b} c_{x b}
        + \Big( v_1 c^\dag_{x a} c_{x b} \\
        & + \frac{v_2}{2} \Big[
            c^\dag_{(x + 1) a} c_{x b} + c^\dag_{(x + 1) b} c_{x a} \Big] \\
        & + \frac{i v_2}{2} \Big[
            c^\dag_{(x + 1) a} c_{x a} - c^\dag_{(x + 1) b} c_{x b} \Big]
        + \mathrm{h.c.} \Big) \bigg\}, \\
    \mathcal{H}^{\mathrm{eb}}_{\mathrm{int}}
    = U_0 \sum_{z = 1}^{2N} & n_z \left( n_z - 1 \right)
        + \sum_{r \geq 1} U_r \left( \sum_{z = 1}^{2N} n_z n_{z + r} \right),
\end{split}\end{equation}
over $N$ unit cells. Above, $v_1, v_2 \geq 0$ are intra- and inter-cell hopping coefficients, $\gamma > 0$ is a loss rate on $b$-sublattices, $U_0 \in \mathbb{R}$ is an on-site interaction strength, and $U_r \in \mathbb{R}$ are range-$r$ density-density interaction strengths between sites on the flattened ladder. The operator $c_{x \ell}$ annihilates a particle at sublattice $\ell \in \{a, b\}$ of unit cell $x \in [N]$, and $\smash{n_z = c^\dag_{x \ell} c_{x \ell}}$ is the number operator with index $z = 2 x - \delta_{\ell a} \in [2N]$ labeling sites on the ladder with sublattice structure flattened. Open boundary conditions (OBC) are imposed by zeroing $c_{x \ell} = 0$ for $x \notin [N]$; periodic boundary conditions (PBC) correspond to setting $c_{x \ell} = c_{(x + N) \ell}$. The Hamiltonian is manifestly non-Hermitian (as $\gamma > 0$). Writing $\smash{\mathcal{H}^{\mathrm{eb}} = \mathcal{H}^{\mathrm{eb}}_{\mathrm{H}} - i \mathcal{H}^{\mathrm{eb}}_{\mathrm{A}}}$ for Hermitian and anti-Hermitian components $\smash{\mathcal{H}^{\mathrm{eb}}_{\mathrm{H}}}$ and $\smash{-i \mathcal{H}^{\mathrm{eb}}_{\mathrm{A}}}$, we have
\begin{equation}\begin{split}
    \mathcal{H}^{\mathrm{eb}}_{\mathrm{A}} = \gamma \sum_{x = 1}^N c^\dag_{x b} c_{x b},
    \label{eq:methods/model/H-edge-burst-anti}
\end{split}\end{equation}
containing the non-Hermitian loss.

The non-interacting $\mathcal{H}^{\mathrm{eb}}_0$ is identical to the paradigmatic lossy quantum walk Hamiltonian studied in Refs.~\cite{xue2022non, yuce2023non}. Notably, the non-Hermitian skin effect (NHSE) is present for all $v_1 \neq 0$, as implied by an equivalence of $\mathcal{H}^{\mathrm{eb}}_0$ to the non-Hermitian Su-Schrieffer-Heeger (SSH) model~\cite{yao2018edge} with left-right asymmetric hoppings under a unitary basis transformation. Alternatively, the NHSE can be understood as arising from the suppression of the $+\vu{x}$-direction portion of flux-induced rotational motion by the $\gamma$ loss, as mentioned in the main text. The imaginary (\ie~dissipative) gap closes~\cite{xue2022non} for $v_1 \leq v_2$, which is the canonical regime where the non-Hermitian edge burst manifests.

In $\mathcal{H}^{\mathrm{eb}}_{\mathrm{int}}$, we take the $U_0 \to \infty$ limit for simplicity, such that hardcore bosonic statistics arise~\cite{kormos2014analytic}. We allow $U_r$ to be independently tuned for each $r$ to access a variety of multi-particle interacting edge burst phenomena. Note that $\mathcal{H}^{\mathrm{eb}}$ possesses a $\mathrm{U}(1)$ number conservation symmetry in the $c$ bosons.

\textbf{Qubit encoding.}
\label{sec:methods/encoding}
To perform simulation of $\mathcal{H}^{\mathrm{eb}}$ on quantum hardware, a mapping between basis states of the system and qubit states must be fixed, which also induces a mapping between the operators of the system ($c_{x \ell}$) and qubit operators. As $\mathcal{H}^{\mathrm{eb}}$ is number-conserving, its Hilbert (Fock) space is the direct sum of disconnected Fock sectors identified by particle numbers. We leverage this structure in state encoding. In the single-particle sector, we identified
\begin{equation}\begin{split}
    \ket{(x, \ell)}_{\mathrm{sys}} = \ket{2 x - \delta_{\ell a} - 1}_{\mathrm{q}},
\end{split}\end{equation}
where $\smash{\ket{(x, \ell)}_{\mathrm{sys}}}$ is the system state with a particle at unit cell $x \in [1, N]$ and sublattice $l \in \{a, b\}$, and $\smash{\ket{2 x - \delta_{\ell a} - 1}_{\mathrm{q}}}$ is the associated computational basis state on the qubits\footnote{Throughout our work an implicit conversion of integer labels for qubit states to binary representation is assumed, that is, $\smash{\ket{z} = \ket{\overline{z}} = \ket{\overline{z}_1} \otimes \ket{\overline{z}_2} \otimes \cdots}$ for $z \in \mathbb{N}$ and $\smash{\overline{z}}$ the binary representation (\textit{i.e.}~bitstring) of $z$, and $\smash{\overline{z}_1}, \smash{\overline{z}_2}, \ldots$ are its digits (bits).}. Thus $\smash{N_{\mathrm{q}} = \ceil{\log_2 2N}}$ system qubits are required to represent the $\mathcal{H}^{\mathrm{eb}}$ quantum ladder in this Fock sector. 

More generally, in the $p$-particle sector, we used a bijection between system states $\smash{\ket{ \{(x_1, \ell_1), \ldots, (x_p, \ell_p)\} }_{\mathrm{sys}}}$, which denote particle occupation at distinct sites $\{(x_1, \ell_1), \ldots, (x_p, \ell_p)\}$ for $x_k \in [1, N]$, $\ell_k \in \{a, b\}$, and qubit states in lexicographic ascending order. Thus minimally $\smash{N_{\mathrm{q}}} = \smash{\ceil{\log_2 (2N)! / (2N - p)!}} \in \order{p \log_2 N}$ qubits are required to represent the quantum ladder. This mapping is similar to first-quantization encoding used in quantum chemistry~\cite{kassal2008polynomial, berry2018improved, babbush2019quantum} as opposed to second-quantization mapping~\cite{babbush2018low}, which requires a number of qubits independent of $p$ but linear in $N$.

\textbf{Fock-sector error detection.}
\label{sec:methods/fock-post-selection}
In our experiments, the number of available qubit states may exceed the number of system states---therefore some qubit states may be unused in state encoding (see \methodsabove{sec:methods/encoding}). These qubit states have no physical meaning and should not be involved in time-evolution of any physical system state. Nonetheless, hardware noise may cause these states to be erroneously populated. We discard circuit shots that measured occupation of these states, as these results are unphysical and indicate hardware error.

\textbf{Time-evolution on non-Hermitian Hamiltonians.} 
\label{sec:methods/time-evolve}
Given a local Hamiltonian $\mathcal{H}$, which need not be Hermitian, an initial state $\rho_0$, and an evolution time $t \geq 0$, we describe an algorithm to perform normalized time-evolution,
\begin{equation}\begin{split}
    \rho_0 \to \rho_t = \mathcal{N}\left[ 
        e^{-i \mathcal{H} t} \rho_0 e^{+i \mathcal{H}^\dag t} \right],
    \label{eq:methods/time-evolve/u-normalized-rte}
\end{split}\end{equation}
where $\mathcal{N}[\rho] = \rho / \tr(\rho)$ normalizes a quantum state. Writing $\mathcal{H} = \mathcal{H}_{\mathrm{H}} - i \mathcal{H}_{\mathrm{A}}$ such that $\mathcal{H}_{\mathrm{H}}$ and $-i \mathcal{H}_{\mathrm{A}}$ are respectively the Hermitian and anti-Hermitian components of $\mathcal{H}$, we split (\ie~trotterize) the evolution into $m$ time steps, such that
\begin{equation}\begin{split}
    \left( \mathcal{M}_{\mathrm{A}} \circ \mathcal{M}_{\mathrm{H}} \right)^m 
    (\rho_0) 
    = \rho_t + \order{\frac{1}{m}},
    \label{eq:methods/time-evolve/time-stepping}
\end{split}\end{equation}
where $\mathcal{M}_{\mathrm{H}}$ and $\mathcal{M}_{\mathrm{A}}$ are quantum maps that we seek to implement, satisfying
\begin{equation}\begin{split}
    \mathcal{M}_{\mathrm{H}}(\rho)
    &= e^{-i \mathcal{H}_{\mathrm{H}} \Delta t} \rho 
        e^{+i \mathcal{H}_{\mathrm{H}} \Delta t}
        + \order{\frac{1}{m^2}}, \\
    \mathcal{M}_{\mathrm{A}}(\rho)
    &= \mathcal{N} \left[ e^{-\mathcal{H}_{\mathrm{A}} \Delta t} \rho
        e^{-\mathcal{H}_{\mathrm{A}} \Delta t} \right] 
        + \order{\frac{1}{m^2}},
    \label{eq:methods/time-evolve/channels}
\end{split}\end{equation}
for the time interval $\Delta t = t / m$ and any arbitrary $\rho$. The error scaling in \cref{eq:methods/time-evolve/time-stepping} implies that arbitrary precision in time-evolution can be achieved by increasing $m$. 

The main difficulty is in realizing $\mathcal{M}_{\mathrm{A}}$ as the transformation is non-unitary. The core idea is to utilize a dilated Hilbert space with an ancillary register, which can be as small as a single qubit, and to implement a joint unitary such that the desired transformation is applied on the system register upon measurement of the ancillae. We detail a systematic non-variational approach for constructing such a circuit with classical processing costs polynomial in the size of the system (see \hyperref[sec:methods/implementation-M-A]{below}). For future use, we denote the Pauli decompositions
\begin{equation}\begin{split}
    \mathcal{H}_{\mathrm{H}} = \sum_{k = 1}^{K_{\mathrm{H}}} \alpha_k \sigma^k,
    \qquad 
    \mathcal{H}_{\mathrm{A}} = \sum_{k = 1}^{K_{\mathrm{A}}} \beta_k \chi^k,
    \label{eq:methods/time-evolve/pauli-decomp}
\end{split}\end{equation}
for $\alpha_k, \beta_k \in \mathbb{R}$ and Pauli strings $\sigma^k, \chi^k$ assumed to be known from the given $\mathcal{H}$. As $\mathcal{H}$ is local, the number of Pauli terms $K_{\mathrm{H}}, K_{\mathrm{A}} \in \poly(N)$.

\textbf{Implementation of $\mathcal{M}_{\mathrm{H}}$.}
\label{sec:methods/implementation-M-H}
The unitary channel $\mathcal{M}_{\mathrm{H}}$ performs time-evolution by the Hermitian Hamiltonian $\mathcal{H}_{\mathrm{H}}$ and can be implemented through any product formula. In our work, we used the first-order Lie-Trotter formula to set
\begin{equation}\begin{split}
    \mathcal{M}_{\mathrm{H}}(\rho) = U_{\mathrm{H}} \rho U_{\mathrm{H}}^\dag, 
    \qquad
    U_{\mathrm{H}} = \prod_{k = 1}^{K_{\mathrm{H}}} e^{-i \alpha_k \sigma^k \Delta t},
    \label{eq:methods/implementation-M-H/trotterization}
\end{split}\end{equation}
where $\{(\alpha_k, \sigma^k)\}_k$ is the Pauli decomposition of $\mathcal{H}_{\mathrm{H}}$ as in \cref{eq:methods/time-evolve/pauli-decomp}. Thus, the circuit for $\mathcal{M}_{\mathrm{H}}$ consists of consecutive layers of exponentiated Pauli strings, each implementable with a standard construction of an $R^z$ rotation on a single qubit sandwiched by CX gates and single-qubit basis changes spanning the support of the Pauli string (see Supplementary Note~\inertlink{2A} for further details). On $n$ qubits with nearest-neighbor qubit connectivity, this standard implementation of an exponentiated Pauli string is $\order{n}$ depth in the worst case.

We utilized several techniques to achieve circuits with fewer CX gates and lower depth than naïve construction. For isolated subsets of Pauli strings of weight $\leq 2$ sharing the same support, we invoked KAK decomposition~\cite{tucci2005introduction}, which produces circuit components for $\mathrm{U}(4)$ unitaries comprising $1$-qubit rotations and at most $3$ CX gates. For subsets of Pauli strings of weight $\leq 3$ sharing the same support, we used the QSEARCH algorithm~\cite{davis2020towards} in the public BQSKit toolkit~\cite{younis2021berkeley} to produce circuit components, which promises optimal-depth synthesis for unitaries up to $4$ qubits. 

For remaining Pauli strings of larger weight, we used a circuit construction procedure that takes advantage of simultaneous diagonalization~\cite{van2020circuit}. We partitioned the Pauli strings into commuting groups by finding a clique cover on their commutation graph\footnote{Determining the \textit{minimum} clique cover, which is an NP-complete problem in the worst case, is not necessary for this method to work, though generally the fewer the number of cliques the smaller the resultant circuit.}. Let $\{P_k\}_k$ be one such commuting group on $n$ qubits; then there exists a Clifford unitary $C$ that simultaneously diagonalizes the group, that is, $C P_k C^\dag = Q_k$ for $Q_k \in \{\mathbb{I}, \sigma^z\}^{\otimes n}$, for all $k$. We solved for $C$ and its circuit implementation, and the diagonal $\{Q_k\}_k$, through a classically efficient stabilizer-based method~\cite{van2020circuit, gokhale2019minimizing}. The circuit for the exponentiated $\{P_k\}_k$ then follows the structure $\smash{C^\dag \left[ \prod_k \exp(-i Q_k \theta) \right] C}$, and suitable orderings of the $Q_k$ terms according to their support allow cascading cancellation of CX gates between neighboring terms~\cite{van2020circuit}. This approach provides reductions in circuit depth and number of CX gates compared to the naïve concatenation of $\exp(-i P_k \theta)$ circuit components in practice.

\textbf{Implementation of $\mathcal{M}_{\mathrm{A}}$.} 
\label{sec:methods/implementation-M-A}
We devised a scheme based on the linear combination of unitaries (LCU) circuit pattern to implement the non-unitary $\smash{e^{-\mathcal{H}_{\mathrm{A}} \Delta t}}$, which is equivalent to imaginary-time evolution under $\smash{\mathcal{H}_{\mathrm{A}}}$. In general, equipped with unitaries $U_0, \ldots, U_{d - 1}$ controlled respectively by the $\ket{0}, \ldots, \ket{d - 1}$ states of an ancillary qubit register, a compact circuit primitive is known~\cite{childs2012hamiltonian, berry2015simulating} that probabilistically implements the action $(\gamma_0 U_0 + \ldots + \gamma_{d - 1} U_{d - 1})$ for coefficients $\gamma_k > 0$ on a system register up to normalization of the resultant quantum state (see \cref{fig:schematic/schematic-circuit-lcu}). This LCU primitive requires $\ceil{\log_2 d}$ ancillary qubits, and comprises an initialization unitary $V$ on the ancillary register dependent on the coefficients, application of the controlled $U_k$ unitaries, and $V^\dag$ followed by measurements on the ancillary register. The primitive succeeds when the ancillae report $\ket{0}$ upon measurement and fails otherwise.

We first establish a general method for an arbitrary Hamiltonian $\mathcal{H}_{\mathrm{A}}$. The central idea is to approximate $e^{-\mathcal{H}_{\mathrm{A}} \Delta t}$ using a linear combination of forward and backward real time-evolution. Specifically, we examine expansions of the type
\begin{equation}\begin{split}
    A_0 \mathbb{I} + \sum_{b = 1}^B A_b \left(
        e^{+i \mathcal{R} \Delta \tau_b} + e^{-i \mathcal{R} \Delta \tau_b} 
    \right)
    = e^{-\mathcal{H}_{\mathrm{A}} \Delta t} + \order{\frac{1}{m^\kappa}},
    \label{eq:methods/implementation-M-A/expansion}
\end{split}\end{equation}
for coefficients $A_0, \ldots, A_B \geq 0$, rescaled simulation times $\{ \Delta \tau_b \}_b$, and a Hermitian auxiliary Hamiltonian $\mathcal{R}$ satisfying $\mathcal{R}^2 = \mathcal{H}_{\mathrm{A}}$. To satisfy error requirements of $\mathcal{M}_{\mathrm{A}}$ as declared in \cref{eq:methods/time-evolve/channels}, we require an approximation order $\kappa \geq 2$. The identity $\mathbb{I}$ and time-evolution propagators $e^{\pm i \mathcal{R} \Delta \tau_b}$ are unitary, thus their linear combination can be implemented by the LCU primitive above. In particular, the time-evolution propagators $e^{\pm i \mathcal{R} \Delta \tau_b}$ can be implemented via Lie-Trotter product formulae, in the same way as $\mathcal{M}_{\mathrm{H}}$, and the ancillae controls for the LCU need only be added to the central $R^z$ rotation for each exponentiated Pauli string. We refer readers to Supplementary Note~\inertlink{2} for elaboration on the lower-level implementation details.

For any $\kappa \geq 2$, solutions for the expansion can be identified by matching Taylor expansion terms by order on both sides of \cref{eq:methods/implementation-M-A/expansion}. In broad families of condensed-matter lattice models, either choices of Hermitian roots $\mathcal{R}$ are known or analytical properties of the Hamiltonians enable $\mathcal{R}$ to be efficiently computed. For example, the roots of topological insulators and superconductors in one and higher dimensions~\cite{ezawa2020systematic, marques2021one, kremer2020square, wu2021square}, and generalized classes of symmetry-protected topological or topologically ordered systems~\cite{deng2022nth, song2022square, marques2021weak, lin2021square, guo2023realization, geng2024quartic} are well-studied. In cases where $\mathcal{R}$ is difficult or inefficient to determine, we provide also an alternative method relying only on knowledge of $\mathcal{H}_{\mathrm{A}}$---see Supplementary Note~\inertlink{2B}.

Minimally, choosing $d = 2$ terms in the LCU necessitates only a single ancillary qubit. This corresponds to a cosine approximation to $e^{-\mathcal{H}_{\mathrm{A}} \Delta t}$ with a single pair $B = 1$ in \cref{eq:methods/implementation-M-A/expansion}, with coefficients $A_0 = 0$ and $A_1 = 1 / 2$, and rescaled simulation times $\Delta \tau = \sqrt{2 \Delta t}$, achieving approximation order $\kappa = 2$. Solutions with larger number of terms $d$ achieving higher approximation order $\kappa$ can readily be found---see Supplementary Table~\inertlink{S6} for examples. 

In the present edge burst context, Hermitian auxiliary Hamiltonians $\mathcal{R}^{\mathrm{eb}}$ satisfying $\smash{(\mathcal{R}^{\mathrm{eb}})^2} = \mathcal{H}_{\mathrm{A}}$ can be found analytically. However, more carefully exploiting the structure of $\mathcal{H}^{\mathrm{eb}}_{\mathrm{A}}$ in fact allows a solution to \cref{eq:methods/implementation-M-A/expansion} with zero approximation error (effectively $\kappa \to \infty$). As $\mathcal{H}^{\mathrm{eb}}_{\mathrm{A}}$ comprises only on-site terms, we find
\begin{equation}\begin{split}
    & e^{-\mathcal{H}^{\mathrm{eb}}_{\mathrm{A}} \Delta t} 
    = \sum_{x = 1}^N \left( n_{x a}
        + e^{-\gamma \Delta t} n_{x b} \right)
    = \frac{1}{2 \eta} \left(U^{\mathrm{eb}}_+ + U^{\mathrm{eb}}_-\right),
\end{split}\end{equation}
for forward and backward time-evolution unitaries $\smash{U^{\mathrm{eb}}_\pm} = \smash{e^{\pm i \mathcal{H}^{\mathrm{eb}}_{\mathrm{aux}}}}$ on the effective Hermitian Hamiltonian
\begin{equation}\begin{split}
    \mathcal{H}^{\mathrm{eb}}_{\mathrm{aux}}
    = \arccos(e^{-\gamma \abs{\Delta t}}) \begin{dcases}
        \sum_{x = 1}^N n_{x b} & \Delta t \geq 0, \\
        \sum_{x = 1}^N n_{x a} & \Delta t < 0,
    \end{dcases}
\end{split}\end{equation}
and scaling factor $\eta = \min(1, e^{\gamma \Delta t})$. Here we present solutions for both $\Delta t \geq 0$ and $\Delta t < 0$ cases for completeness. Thus an exact LCU construction to implement $\smash{e^{-\mathcal{H}^{\mathrm{eb}}_{\mathrm{A}} \Delta t}}$ is a priori known. We use this solution in our experiments, with $U^{\mathrm{eb}}_\pm$ implemented via first-order Trotter-Lie product formula.

\textbf{Qubit reset and re-use.}
\label{sec:methods/resets}
We performed mid-circuit reset of the ancillary qubit, used in the LCU circuit implementation of the $\mathcal{M}_{\mathrm{A}}$ map (see \methodsabove{sec:methods/implementation-M-A}), to the $\ket{0}$ state after each time step such that the same qubit is re-used over the entire time-evolution experiment. Each reset is performed by executing an $X$ gate on the qubit conditioned on the outcome of the measurement in the preceding $\mathcal{M}_{\mathrm{A}}$ circuit component. The mid-circuit readout and feedforward capability of the quantum processors~\cite{baumer2024quantum, baumer2023efficient, corcoles2021exploiting} enable this resource-saving measure.

\textbf{Quantum state normalization.}
\label{sec:methods/normalization}
Quantum states on a quantum computer are normalized by nature of the platform. Above, we described the implementation of normalized time-evolution of an initial state [\cref{eq:methods/time-evolve/u-normalized-rte}]. However, in some simulation settings, as in the present edge burst context, the process of interest is time-evolution without normalization,
\begin{equation}\begin{split}
    \omega_0 \to \omega_t = e^{-i \mathcal{H} t} \omega_0 e^{+i \mathcal{H}^\dag t},
    \label{eq:methods/normalized/v-normalized-rte}
\end{split}\end{equation}
where $\omega_0 = \rho_0$ is a normalized initial state but $\omega_t$ is not normalized, $\tr(\omega_t) \neq 1$. Enforcing normalization $\mathcal{N}(\omega_t) = \rho_t$ produces the normalized time-evolution in \cref{eq:methods/time-evolve/u-normalized-rte}, which is realized on the quantum platform. Here we examine the recovery of $\omega_t$ from $\rho_t$. In particular, we describe two methods of recovering the quantum state normalization factor $\smash{A_t = \sqrt{\tr(\omega_t)} > 0}$ such that $\omega_t = A_t^2 \rho_t$. 

The first method is general for any lossy $\mathcal{H}$ and $t \geq 0$ such that $\smash{\norm{e^{-i \mathcal{H} t}}} \leq 1$, and works by examining the success probability $S_t$ of the normalized time-evolution algorithm (as detailed \hyperref[sec:methods/time-evolve]{above}). In each time step, failure of the algorithm can occur, which is detected by a $\ket{1}$ measurement outcome on the ancillary qubit, and indicates an incorrect projection of the quantum state that realizes a complementary action $\mathbb{I} - e^{-\mathcal{H}_{\mathrm{A}} \Delta t}$ instead of the desired $e^{-\mathcal{H}_{\mathrm{A}} \Delta t}$. The decrease of quantum state norm over time directly manifests in the probability of success, in particular, $A_t^2 \approx S_t$. Thus $S_t$ measured in experiments directly allow recovery of $A_t$, up to errors in circuit construction (\eg~trotterization and LCU approximation) and hardware noise that distort the observed $S_t$. See Supplementary Note~\inertlink{4C} for technical elaboration and experimental investigation.

The second method is through time-integration of site-resolved occupancy data, as described in the main text. As $\mathcal{H}^{\mathrm{eb}}$ is lossy on $b$-sublattices, we obtain
\begin{equation}\begin{split}
    \dv{t} \tr(\omega_t) 
    &= -2 \expval{\mathcal{H}^{\mathrm{eb}}_{\mathrm{A}}}_{\omega_t}
    = -2 \gamma A_t^2 \sum_{x = 1}^N \expval{n_{x b}}_{\rho_t },
\end{split}\end{equation}
where $\smash{\expval{n_{x b}}_{\rho_t} = \tr\left( \rho_t n_{x b} \right)}$ is the occupancy of sublattice $b$ of unit cell $x$ as measured on the $\rho_t$ state on the quantum platform. This is a restatement of \cref{eq:results/model/wavefunction-norm} of the main text. At the same time, by definition of $A_t$, $(\dd / \dd t) \tr(\omega_t) = (\dd / \dd t) A_t^2 \tr(\rho_t) = 2 A_t (\dd A_t / \dd t)$. Thus we have
\begin{equation}\begin{split}
    \dv{A_t}{t} = - \gamma A_t \sum_{x = 1}^N \expval{n_{x b}}_{\rho_t},
\end{split}\end{equation}
with the solution for $A_t$ reported in \cref{eq:results/simulation/norm-time-integration}. Thus $A_t$ can be recovered through numerical integration of $b$-sublattice occupancies measured on normalized time-evolved states in experiments, sampled over the time domain $[0, t]$. This is the method used in all our experiments discussed in the main text (\cref{fig:results-1p-small-system,fig:results-1p-large-system,fig:results-1p-spectral,fig:results-2p,fig:results-cluster}) as well as Supplementary Figures~\inertlink{S2} to \inertlink{S5}.

\textbf{Escape probabilities.}
\label{sec:methods/escape-probs}
As $\mathcal{H}^{\mathrm{eb}}$ is lossy on $b$-sublattices, as reported in \cref{eq:results/model/escape-probs}, the probability of escape from unit cell $x$ by time $t$ is given by
\begin{equation}\begin{split}
    P_x(t) 
    = 2 \gamma \int_0^t \expval{n_{x b}}_{\omega_\tau} \, \dd{\tau}
    = 2 \gamma \int_0^t A_\tau^2 \expval{n_{x b}}_{\rho_\tau} \, \dd{\tau},
\end{split}\end{equation}
where $A_\tau$ are quantum state normalization factors recovered from experiment data and $\expval{n_{x b}}_{\rho_\tau}$ are $b$-sublattice occupancies on normalized time-evolved states measured in experiments. The total escape probability across all unit cells is $\smash{P(t) = \sum_{x = 1}^N P_x(t)}$ accordingly. Note $P_x(0) = P(0) = 0$, and that $P_x(t)$ and $P(t)$ are non-decreasing with $t$. In the long-time limit, each $P_x(t)$ approaches an asymptotic (\textit{i.e.}~steady-state) value $\mathcal{P}_x = \lim_{t \to \infty} P_x(t)$, and $P(\infty) = 1$ as expected of a properly normalized probability distribution, consistent with Ref.~\cite{xue2022non}. 

The dynamics of the system tends to completion as $P(t) \to 1$, and there is no further purpose in assessing dynamics at larger times which contribute negligibly to escape probabilities. We terminate time-evolution at $t$ sufficiently large to achieve $P(t) \gtrsim 0.995$ in our experiments, more than sufficient to clearly observe edge burst phenomenology as demonstrated.

\textbf{Measuring imaginary energy gap.}
\label{sec:methods/max-min-imag-E}
Time-evolution by $\mathcal{H}$ as written in \cref{eq:methods/time-evolve/u-normalized-rte} encodes an effective imaginary-time evolution with respect to the anti-Hermitian component of $\mathcal{H}$ in the following sense. Suppose $\ket{\phi}$ is a (right) eigenstate of $\mathcal{H}$ with eigenenergy $E \in \mathbb{C}$, that is, $\mathcal{H} \ket{\phi} = E \ket{\phi}$. Then
\begin{equation}\begin{split}
    e^{-i \mathcal{H} t} \ket{\phi}
    &= e^{-i E t} \ket{\phi}
    = e^{-i \Re(E) t} e^{\Im(E) t} \ket{\phi}, \\
    \frac{\norm{e^{-i \mathcal{H} t} \ket{\phi}}}{\norm{\ket{\phi}}}
    &= e^{\Im(E) t}.
\end{split}\end{equation}

Let us consider the time-evolution of an initial state $\rho_0$ which comprises a mixture of eigenstates of $\mathcal{H}$. Then for $t > 0$ ($t < 0$), the amplitudes of all other eigenstates become exponentially damped compared to the eigenstate with the largest (smallest) $\Im(E)$. Thus, for sufficiently large $\abs{t}$, the time-evolved state $\rho_t$ purifies to an eigenstate of extremal $\Im(E)$; the required $\abs{t}$ can be estimated by demanding that $P(t)$ converges close to unity (see \hyperref[sec:methods/escape-probs]{above}). After time-evolution, the eigenenergy $E$ can be measured through standard techniques. We used Hamiltonian averaging~\cite{mcclean2016theory, mcclean2014exploiting, peruzzo2014variational} in our experiments,
\begin{equation}\begin{split}
    \Im(E) &= -\expval{\mathcal{H}_{\mathrm{A}}}_{\rho_t}
    = -\sum_{k = 1}^{K_{\mathrm{A}}} \beta_k \expval*{\chi^{(k)}}_{\rho_t},
\end{split}\end{equation}
where as before $\smash{\expval{O}_{\rho_t} = \tr(\rho_t O)}$ is the measured expectation value of an observable $O$ on $\rho_t$ on the quantum device. As $\smash{\chi^{(k)}}$ are Pauli strings, measuring their expectation values on a quantum processor involves only Clifford basis changes before computational-basis readouts. To reduce the number of circuits required, commuting subsets of $\smash{\{\chi^{(k)}\}_k}$ can be measured simultaneously on the same circuit, with the joint Clifford basis change given by an efficient stabilizer-based method~\cite{gokhale2019minimizing}. 

We are interested in the imaginary energy (\textit{i.e.}~dissipative) gap in the present work, which is open when the largest $\Im(E) < 0$ and closed otherwise. We took the initial state $\rho_0 = \mathcal{N}(\mathbb{I})$ to be the maximally mixed state, equivalently the infinite-temperature Gibbs state, prepared through a single round of mid-circuit computational-basis measurements on system qubits initialized in $\ket{+}$ states, whose outcomes are discarded (see Supplementary Figure~\inertlink{S1} for circuit schematic).

\textbf{Circuit recompilation.}
\label{sec:methods/recompilation}
We follow the formulation established in prior works, in particular Refs.~\cite{koh2022stabilizing, koh2022simulation, koh2024realization, sun2021quantum, khatri2019quantum, heya2018variational, conlon2025attainability}. Circuit recompilation is performed using a circuit ansatz whose parameters are dynamically optimized. We use an ansatz comprising an initial layer of single-qubit rotations ($U_3$ gates) on all qubits followed by $K \leq 20$ ansatz layers, each comprising a layer of CX gates entangling adjacent qubits and a layer of $U_3$ rotations (see Supplementary Figure~\inertlink{S1}). Each $U_3$ gate in the ansatz is associated with rotation angles $(\theta, \phi, \lambda)$; we collate them into a parameter vector $\vb*{\vartheta}$. Then, given a target circuit component unitary $V$ and an initial state $\ket{\psi_0}$, the optimization problem
\begin{equation}\begin{split}
    \argmax_{\vb*{\vartheta}} \mathcal{F}(V_{\vb*{\vartheta}} \ket{\psi_0}, V \ket{\psi_0})
    = \argmax_{\vb*{\vartheta}} \abs{\mel{\psi_0}{V_{\vb*{\vartheta}}^\dagger V}{\psi_0}}^2,
\end{split}\end{equation}
is numerically treated, where $V_{\vb*{\vartheta}}$ is the circuit ansatz unitary with parameters $\vb*{\vartheta}$. The recompiled circuit component is then the ansatz with optimal parameters fixed. As $\mathcal{H}^{\mathrm{eb}}$ is number-conserving, to enhance recompilation performance and the quality of recompiled circuits, we focus optimization on the Fock sectors relevant to the time-evolution simulation~\cite{koh2024realization}. In our implementation, we estimate $V_{\vb*{\theta}}$ through auto-differentiable tensor network-based ansatz simulation~\cite{gray2018quimb}, and use L-BFGS-B with basin-hopping to perform the optimization~\cite{andrew2007scalable}. Recompilation is set to never exceed a few minutes per circuit.

\textbf{Readout error mitigation.} 
\label{sec:methods/readout-mitigation}
The quantum devices used in our experiments present non-negligible probabilities (${\sim} 1$--$3\%$) of reporting a $\ket{1}$ outcome when the measured qubit is in $\ket{0}$ and vice versa, typical for current-era quantum hardware. This is referred to as readout error and is caused, for example, by ambiguity in microwave signal characteristics (\ie~IQ data) that inhibit classification of measurement outcomes. We mitigated the effect of these errors by characterizing the measurement bit-flip probabilities of the qubits used, then performing linear inversion to approximately recover the true measurement counts from observed measurement counts~\cite{kandala2017hardware, kandala2019error, jurcevic2021demonstration}. Unlike prior works~\cite{kandala2017hardware, kandala2019error, jurcevic2021demonstration, koh2022stabilizing, koh2022simulation, koh2024realization}, our experiment circuits contain mid-circuit measurements---in particular a qubit can be measured multiple times throughout the circuit. We thus employed a modified scheme as described below.

Consider an experiment circuit containing measurements on $n$ qubits, labeled by $[n]$, containing $L$ layers of measurements. The set of qubits measured in layer $l$ is $\mathcal{I}_l \subseteq [n]$, of size $p_l = \abs*{\mathcal{I}_l}$, and the total number of measurements over all layers is $p \geq n$. The measurement outcomes of each shot of the circuit form a bitstring $x \in \{0, 1\}^p$, where entry $x_j$ records the outcome of the $j^{\text{th}}$ measurement in the circuit. Upon completion of a specified number of shots, a raw counts vector $\vb{c}$ is returned, where entry $c_x$ records the number of times the bitstring $x$ was observed.

We assume that there are no correlations in readout errors for measurements in different layers as these measurements are separated in time, consistent with standard Markovian assumptions on noise channels~\cite{suter2016colloquium, clerk2010intro}; but within the same layer we accommodate arbitrary correlations. Let $M$ be the readout calibration matrix on the $n$ qubits, where each entry $M_{s t}$ records the probability of obtaining outcome $s \in \{0, 1\}^n$ upon measurement of the qubits when the true outcome is $t \in \{0, 1\}^n$. Then $\vb{c}$ and the ideal counts $\smash{\widetilde{\vb{c}}}$ that would hypothetically be observed without readout errors are related by
\begin{equation}\begin{split}
    \vb{c} = \left( \bigotimes_{l = 1}^L 
        M[\mathcal{I}_l] \right)
    \widetilde{\vb{c}}
    \Longleftrightarrow 
    \widetilde{\vb{c}}
    = \left( \bigotimes_{l = 1}^L 
        M[\mathcal{I}_l]^{-1} \right)
    \vb{c},
    \label{eq:methods/readout-mitigation/counts-relation}
\end{split}\end{equation}
where $M[\mathcal{I}_l]$ is $M$ with support retained over the qubits $\mathcal{I}_l$ of layer $l$ but marginalized over the remaining qubits, that is, 
\begin{equation}\begin{split}
    M[\mathcal{I}_l]_{s t} = \frac{1}{2^{n - p_l}}
        \sum_{\substack{s' \in \{0, 1\}^n \\ s'[\mathcal{I}_l] = s}} \,\,
        \sum_{\substack{t' \in \{0, 1\}^n \\ t'[\mathcal{I}_l] = t}} \,\,
        M_{s' t'},
\end{split}\end{equation}
where $s[\mathcal{I}_l]$ denotes the substring formed by entries of a bitstring $s$ at qubit indices $\mathcal{I}_l$. Thus \cref{eq:methods/readout-mitigation/counts-relation} enables the recovery of ideal counts $\smash{\widetilde{\vb{c}}}$ from raw counts $\vb{c}$. To ensure that the mitigated counts $\smash{\widetilde{\vb{c}}}$ are physical, we employed an additional step of projecting $\smash{\widetilde{\vb{c}}}$ onto the nearest non-negative probability distribution (in $l_2$-norm)~\cite{smolin2012efficient}.

The matrix $M$ can be characterized by running calibration circuits, which prepare the $n$ qubits in state $\ket{t}$ for every $t \in \{0, 1\}^n$ and record probabilities of observing each outcome $s \in \{0, 1\}^n$ upon measurement. However, this direct method requires $2^n$ calibration circuits and is prohibitively costly for large $n$. We instead employed a tensored approach~\cite{koh2022stabilizing, koh2022simulation, koh2024realization}, which splits the $n$ qubits into sub-registers containing $\{n_g\}_g$ qubits and assumes that readout errors are uncorrelated between sub-registers. The calibration is performed for each sub-register, producing matrices $\{M^{(g)}\}_g$, and the overall $M = \smash{\bigotimes_g M^{(g)}} \Longleftrightarrow M^{-1} = \smash{\bigotimes_g {M^{(g)}}^{-1}}$. Then, performing linear inversion as in \cref{eq:methods/readout-mitigation/counts-relation}, each $\smash{M^{(g)}}^{-1}$ can act sector-wise on $\vb{c}$, avoiding the cost of assembling the entire $M^{-1}$. Moreover, the calibration circuits for different sub-registers can be merged as they act on disjoint sets of qubits; thus this approach requires $2^{\max_g n_g} \ll 2^n$ calibration circuits. We chose sub-registers such that $n_g \leq 5$ in our experiments. 

While the employed readout error mitigation addresses errors in the measurement outcome statistics, it does not address errors occurring in feedforward---for example, a faulty mid-circuit measurement outcome on an ancillary qubit in our circuits can cause an incorrect reset (see \methodsabove{sec:methods/resets}). A recent work by several of the authors presented a protocol that overcomes this limitation~\cite{koh2024readout} but some experiments in this study predate this development; we do not use this protocol here.

\textbf{Dynamical decoupling.}
\label{sec:methods/dynamical-decoupling}
Idle periods on qubits are invariably present in our experiment circuits, as a qubit may need to wait for gates to complete on other qubits before being involved in $2$-qubit operations with them. During idle periods, qubits are still subject to decoherence (\eg~dephasing noise and thermal relaxation). Dynamical decoupling~\cite{viola1999dynamical}, which runs through alternating basis changes that multiply to the identity, can be used to suppress the effects of such decoherence. At a basic level, dynamical decoupling refocuses the dephasing experienced by the qubits such that the accumulated noise contributions cancel to a certain order.

We employed the common XY4 pulse sequence~\cite{viola1999dynamical, pokharel2018demonstration} for dynamical decoupling in our experiments. The sequence is inserted for idle periods not exceeding the equivalent of $2$ CX gate durations (${\sim}800$--$\SI{1000}{\nano\second}$); for longer idle durations repetitions of the sequence are inserted. While there exist more sophisticated dynamical decoupling sequences that cancel noise to higher orders~\cite{khodjasteh2005fault, uhrig2007keeping}, empirical results~\cite{ezzell2023dynamical} suggest that simple sequences (\eg~XY4) can be more performant in practice, due to the presence of errors on the gates introduced by the sequence and differences between hardware noise backgrounds and theoretical assumptions.

\textbf{Pauli twirling.} 
\label{sec:methods/pauli-twirling}
We applied randomized twirling of CXs on each experiment circuit. Each CX gate in the circuit is replaced by $P_1 (\mathrm{CX}) P_2$, where $(P_1, P_2)$ are randomly drawn Paulis such that the action of the twirled CX remains invariant. See Supplementary Table~\inertlink{S5} for a complete list of $(P_1, P_2)$ pairs used. In general, the twirling of gates in circuits, also referred to as randomized compiling in literature~\cite{hashim2021randomized, wallman2016noise}, has the effect of converting coherent and non-Markovian error sources into stochastic errors which can be averaged over in an experiment. Pauli twirling, in particular, converts error processes into diagonal Pauli channels~\cite{kim2023scalable}. We used Pauli twirling in conjunction with zero-noise extrapolation.

\textbf{Zero-noise extrapolation.} 
\label{sec:methods/zne}
A general method for mitigating inaccuracy of expectation values due to hardware noise is zero-noise extrapolation (ZNE), which functions by collecting additional data at amplified noise levels and then extrapolating into the zero-noise limit~\cite{kim2023evidence, kim2023scalable, majumdar2023best, giurgica2020digital, temme2017error}. We employ a gate-level implementation of ZNE in our experiments; alternative pulse-level methods have been explored in the literature~\cite{garmon2020benchmarking, kandala2019error}. Our procedure was as follows. For an experiment measuring a set of observables $\{O_k\}_k$, we perform repetitions at noise amplification factors $1 = \lambda_1 < \lambda_2 < \ldots < \lambda_r$, where $\lambda = 1$ corresponds to the unmodified experiment. Noise in an experiment is amplified by randomized local gate folding~\cite{giurgica2020digital}, that is, each gate $G$ in each circuit is probabilistically replaced by $G G^\dag G$, such that the total number of 1- and 2-qubit gates in the circuit is each increased by factor $\lambda > 1$. Note that $G^\dag$ is known for all basis gates (\textit{e.g.}~the Paulis and CX are involutory), so compilation of the folded circuits is straightforward. To illustrate, $\lambda = 3$ folds every gate in each circuit, $1 < \lambda < 3$ contains a mixture of folded and unfolded gates, and $\lambda > 3$ can be reached by repeated folding. 

Let $\smash{\expval*{O_k^{(\lambda)}}}$ denote $\expval{O_k}$ measured on hardware at noise amplification factor $\lambda$. We assume an affine relation between $\smash{\expval*{O_k^{(\lambda)}}}$ and $\lambda$, that is, $\smash{\expval*{O_k^{(\lambda)}}} \approx m_k \lambda + c_k$ for gradient $m_k$ and zero-noise intercept $c_k$, which enables a simple formulation of the extrapolation problem. This linearization is valid for $\lambda$ not too large, such that hardware noise does not saturate measured observables. We treat the least-squares regression
\begin{equation}\begin{split}
    (\vb{m}^*, \vb{c}^*) = \argmin_{\vb{m}, \vb{c}} \sum_k \sum_{\gamma = 1}^r
        \left( m_k \lambda_\gamma + c_k - \expval{O_k^{(\lambda_\gamma)}} \right)^2,
\end{split}\end{equation}
and the zero-noise estimates $\smash{\{\expval*{O_k^{(0)}}\}_k}$ are then given by the intercepts $\{c^*_k\}_k$. We additionally impose constraints on $(\vb{m}$, $\vb{c})$ based on $\{O_k\}_k$ to ensure physicality of regression results. In particular, for experiments in the $p$-particle Fock sector measuring site-resolved occupancies, that is $\{O_k\}_k = \{n_{x \ell}\}_{x \ell}$, we imposed number conservation and hardcore bosonic statistics, such that $0 \leq \expval{n_{x \ell}} \leq 1$ for all sites and the sum of occupancies equaled $p$, at all noise levels on the regression lines. For experiments measuring the imaginary energy in the $p$-particle sector, that is $\{O_k\}_k = \{{\mathcal{H}^{\mathrm{eb}}_{\mathrm{A}}}\}$, we imposed $\smash{\expval{\mathcal{H}^{\mathrm{eb}}_{\mathrm{A}}} \leq 0}$ as $\smash{\mathcal{H}^{\mathrm{eb}}}$ is lossy and $-p \gamma \leq \smash{\expval{\mathcal{H}^{\mathrm{eb}}_{\mathrm{A}}}}$ by the variational principle. These constraints translate into conditions on $(\vb{m}$, $\vb{c})$---see Supplementary Note~\inertlink{3}.

The constrained regression is convex and can be solved using standard techniques with guaranteed optimal solutions~\cite{diamond2016cvxpy}. In our experiments, we chose $(\lambda_1, \ldots, \lambda_r) = (1, 1.25, 1.5, 1.75, 2)$ and used the general-purpose operator-splitting quadratic program (OSQP) solver~\cite{stellato2020osqp} to perform the constrained regressions. For each experiment circuit at each $\lambda$, we produced $16$ gate-folded circuits, each further compiled with an independent instance of randomized Pauli twirling (see \methodsabove{sec:methods/pauli-twirling}). Each set of $16$ circuits were executed on hardware and their results averaged, to produce the $\smash{\expval*{O_k^{(\lambda)}}}$ expectation value for regression. See Supplementary Note~\inertlink{3} for further technical details and discussion.

\textbf{Qubit selection.}
\label{sec:methods/qubit-selection}
The quantum devices we utilized have more qubits than needed for our experiments, and there is variation in qubit error rates on each device (see Supplementary Table~\inertlink{S4} for device performance characteristics). We hence used a search procedure to select qubits of the lowest estimated error to use in our simulations, following the demonstrated methods in Refs.~\cite{koh2024realization, koh2023measurement, koh2022simulation, koh2022stabilizing, nation2023suppressing, niu2020hardware}. Given an experiment circuit $\mathcal{C}$ requiring $n$ qubits and the connectivity graph of 2-qubit couplings (CX, ECR, or CZ) on the device, we identified all distinct qubit chains $\mathcal{X}$ of length $n$ on the device, and for each chain $x \in \mathcal{X}$ computed an estimated circuit error $\mathcal{E}_x[\mathcal{C}]$ by summing the calibrated gate errors of each gate in $\mathcal{C}$. We selected the qubit chain $x$ that minimized $\mathcal{E}_x[\mathcal{C}]$. This selection procedure is approximate, as the gate error rates are obtained from routine (${\sim}$daily) calibration of the quantum devices and are subject to drift over time, and the circuit error estimation $\mathcal{E}_x[\mathcal{C}]$ is not exact.

\clearpage
\pagebreak


\foreach \x in {1,...,23}
{%
\clearpage
\includepdf[pages={\x}]{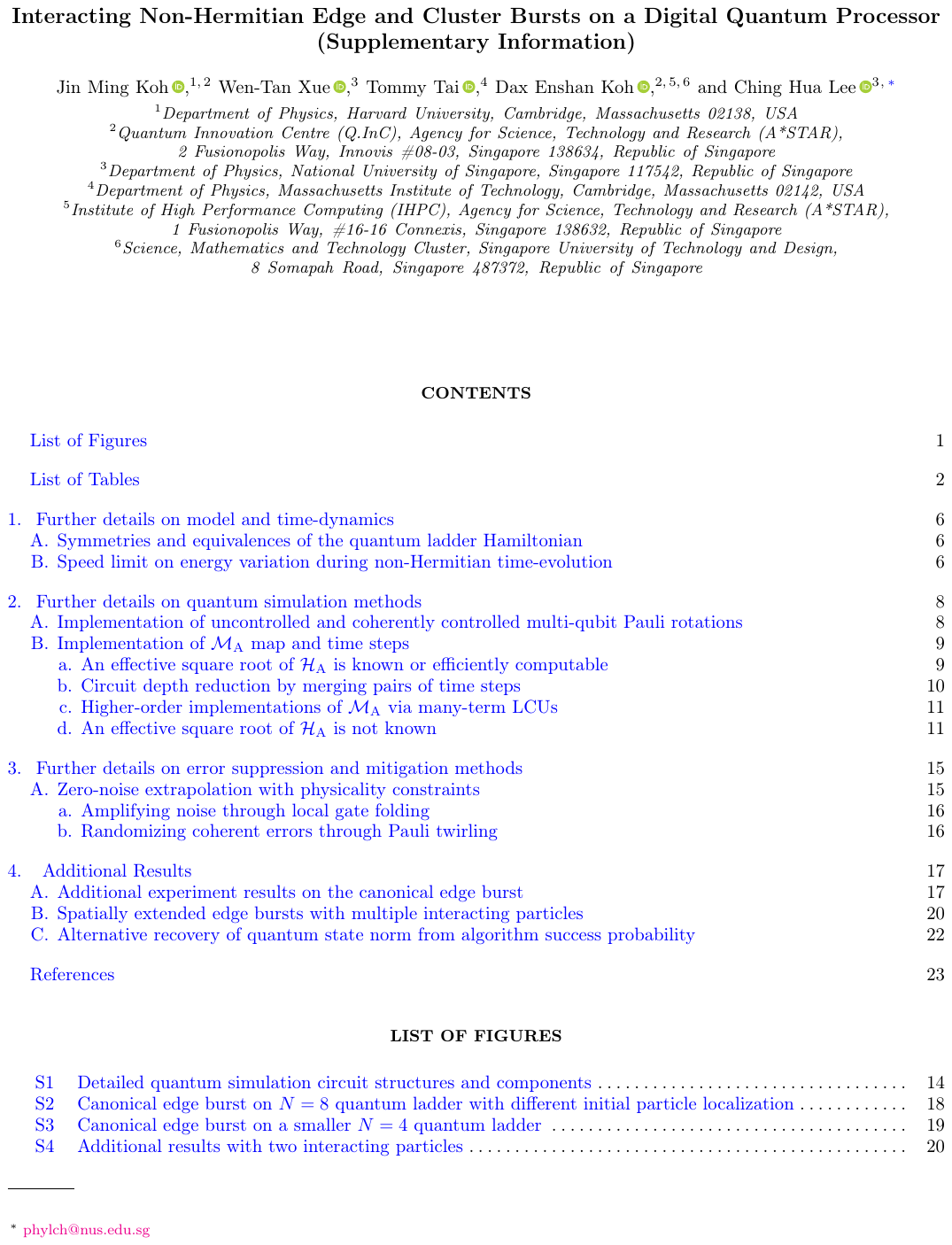}
}

\end{document}